\documentclass[letterpaper,twocolumn,10pt]{article}
\usepackage{usenix,epsfig,endnotes}
\usepackage[utf8]{inputenc}
\usepackage[cmex10]{amsmath}
\usepackage{array}
\usepackage[hyphens]{url}
\usepackage[table]{xcolor}
\usepackage{xspace}
\PassOptionsToPackage{hyphens}{url}\usepackage{url}
\usepackage{tikz}
\usepackage{pgfplots}
\usetikzlibrary{pgfplots.groupplots}
\usetikzlibrary{arrows}
\usetikzlibrary{patterns}
\usetikzlibrary{positioning}
\usetikzlibrary{decorations.pathreplacing}
\usetikzlibrary{shapes.arrows}
\usetikzlibrary{pgfplots.groupplots}
\usepackage{multirow}
\pgfplotsset{compat=newest}
\usepackage{letltxmacro}
\usetikzlibrary{external}
\usepackage{cite}
\usepackage{ifthenx}
\usepackage{collcell}
\usepackage{nicefrac}
\usepackage{booktabs}
\usepackage{amsmath}
\usepackage{filecontents}
\usepackage{listings}
\usepackage{lipsum}
\usepackage{todonotes}
\usepackage{makecell}
\usepackage{paralist}
\usepackage{flushend}
\usepackage{multirow}
\usepackage{caption}
\usepackage{subcaption}
\usepackage{upgreek}
\usepackage{arydshln}
\usepackage{threeparttable}

\LetLtxMacro{\oldtodo}{\todo}
\renewcommand{\todo}[2][]{\oldtodo[#1]{#2}}
\renewcommand{\todo}[1]{\oldtodo[fancyline]{#1}}
\renewcommand{\todo}[2][]{\oldtodo[fancyline,size=\footnotesize,#1]{#2}}
\renewcommand{\todo}[1]{\oldtodo[fancyline,size=\footnotesize]{#1}}

\usetikzlibrary{patterns}
\usepackage{algorithmic}
\usepackage[ruled]{algorithm2e}

\usepackage{color}
\usepackage{paralist}
\lstdefinestyle{mystyle}{
    commentstyle=\color{green!40!black},
    basicstyle=\footnotesize,
    breakatwhitespace=false,
    breaklines=true,
    captionpos=b,
    keepspaces=true,
    numbers=left,
    numbersep=5pt,
    showspaces=false,
    showstringspaces=false,
    showtabs=false,
    tabsize=2,
    xleftmargin=15pt
}

\def\subheading#1{\medskip\noindent{\boldmath\textbf{#1}}~\ignorespaces}

\pgfdeclarepatternformonly{mydots}{\pgfqpoint{-2pt}{-2pt}}{\pgfqpoint{4pt}{4pt}}{\pgfqpoint{2pt}{2pt}}
{
  \pgfsetlinewidth{0.4pt}
  \pgfpathmoveto{\pgfqpoint{0pt}{0pt}}
  \pgfpathlineto{\pgfqpoint{0pt}{3.1pt}}
  \pgfpathmoveto{\pgfqpoint{0pt}{0pt}}
  \pgfpathlineto{\pgfqpoint{3.1pt}{0pt}}
  \pgfusepath{stroke}
}

\newcommand{\etal}{et~al.}
\newcommand{\ie}{\textit{i.e.}, }
\newcommand{\eg}{e.g., }
\mathchardef\mhyphen="2D

\begin{document}

\title{DRAMA: Exploiting DRAM Addressing for Cross-CPU Attacks}

\author{
{\rm Peter Pessl}, {\rm Daniel Gruss}, {\rm Clémentine Maurice}, {\rm Michael Schwarz} and {\rm Stefan Mangard}\\
Graz University of Technology, Austria 
}

\maketitle

\begin{abstract}
  
In cloud computing environments, multiple tenants are often co-located on the same multi-processor system. Thus, preventing information leakage between tenants is crucial.
While the hypervisor enforces software isolation, shared hardware, such as the CPU cache or memory bus, can leak sensitive information.
For security reasons, shared memory between tenants is typically disabled. Furthermore, tenants often do not share a physical CPU.
In this setting, cache attacks do not work and only a slow cross-CPU covert channel over the memory bus is known.
In contrast, we demonstrate a high-speed covert channel as well as the first side-channel attack working across processors and without any shared memory. To build these attacks, we use the undocumented DRAM address mappings.

We present two methods to reverse engineer the mapping of memory addresses to DRAM channels, ranks, and banks. One uses physical probing of the memory bus, the other runs entirely in software and is fully automated. 
Using this mapping, we introduce DRAMA attacks, a novel class of attacks that exploit the DRAM row buffer that is shared, even in multi-processor systems.
Thus, our attacks work in the most restrictive environments.
First, we build a covert channel with a capacity of up to 2\,Mbps, which is three to four orders of magnitude faster than memory-bus-based channels. 
Second, we build a side-channel template attack that can automatically locate and monitor memory accesses.
Third, we show how using the DRAM mappings improves existing attacks and in particular enables practical Rowhammer attacks on DDR4.

\end{abstract}

\section{Introduction}
\bgroup
\let\thefootnote\relax\footnotetext{Original publication in the Proceedings of the 25th Annual USENIX Security Symposium (USENIX Security 2016).\\\url{https://www.usenix.org/conference/usenixsecurity16/technical-sessions/presentation/pessl}}
\egroup

Due to the popularity of cloud services, multiple tenants sharing the same physical server through different virtual machines (VMs) is now a common situation.
In such settings, a major requirement is that no sensitive information is leaked between tenants, therefore proper isolation mechanisms are crucial to the security of these environments. While software isolation is enforced by hypervisors, shared hardware presents risks of information leakage between tenants.
Previous research shows that microarchitectural attacks can leak secret information of victim processes, \eg by clever analysis of data-dependent timing differences.
Such side-channel measurements allow the extraction of secret information like cryptographic keys or enable communication over isolation boundaries via covert channels.

Cloud providers can deploy different hardware configurations, however multi-processor systems are becoming ubiquitous due to their numerous advantages.
They offer high peak performance for parallelized tasks while enabling sharing of other hardware resources such as the DRAM. 
They also simplify load balancing while still keeping the area and cost footprint low.
Additionally, cloud providers now commonly disable memory deduplication between VMs for security reasons.

To attack such configurations, successful and practical attacks must comply with the following requirements:
\begin{compactenum}
  \item \textit{Work across processors:} As these configurations are now ubiquitous, an attack that does not work across processors is severely limited and can be trivially mitigated by exclusively assigning processors to tenants or via the scheduler.
  \item \textit{Work without any shared memory:} With memory deduplication disabled, shared memory is not available between VMs. All attacks that require shared memory are thus completely mitigated in cross-VM settings with such configurations.
\end{compactenum}

In the last years, the most prominent and well-studied example of shared-hardware exploits is cache attacks. They use the processor-integrated cache and were shown to be effective in a multitude of settings, such as cross-VM key-recovery attacks~\cite{Ristenpart2009,Irazoqui2014RAID,Zhang2012,Inci2015iacr898}, including attacks across cores~\cite{Yarom2014Usenix,Liu2015SP,Maurice2015DIMVA,Gruss2016Flush}. However, due to the cache being local to the processor, these attacks do not work across processors and thus violate requirement~1. Note that in a recent concurrent work, Irazoqui~\etal\cite{Irazoqui2015cpca} presented a cross-CPU cache attack which exploits cache coherency mechanisms in multi-processor systems. However, their approach requires shared memory and thus violates requirement~2. 
The whole class of cache attacks is therefore not applicable in multi-processor systems without any shared memory.

Other attacks leverage the main memory that is a shared resource even in multi-processor systems. Xiao~\etal\cite{Xiao2013} presented a covert channel that exploits memory deduplication. This covert channel has a low capacity and requires the availability of shared memory, thus violating requirement~2.
Wu~\etal\cite{Wu2014} presented a covert channel exploiting the locking mechanism of the memory bus. While this attack works across processors, the capacity of the covert channel is orders of magnitude lower than that of current cache covert channels.

Therefore, only a low capacity covert channel and no side-channel have been showed with the two aforementioned requirements so far. In contrast, we demonstrate two attacks that do not use shared memory and work across processors: a high-speed covert channel as well as the first side-channel attack.

\subheading{Contributions.}
Our attacks require knowledge of the undocumented mapping of memory addresses to DRAM channels, ranks, and banks. We therefore present two methods to reverse engineer this mapping.
The first method retrieves the correct addressing functions by performing physical probing of the memory bus. 
The second method is entirely software-based, fully automatic, and relies only on timing differences.\footnote{The source code of this reverse-engineering tool and exemplary DRAMA attacks can be found at \url{https://github.com/IAIK/drama}.}
Thus, it can be executed remotely and enables finding DRAM address mappings even in VMs in the cloud. 
We reverse engineered the addressing functions on a variety of processors and memory configurations. 
Besides consumer-grade PCs, we also analyzed a dual-CPU server system -- similar to those found in cloud setups -- and multiple recent smartphones.

Using this reverse-engineered mapping, we present \textit{DRAMA} attacks, a novel class of attacks that exploit the \textit{DRAM Addressing}.
In particular, they leverage DRAM row buffers that are a shared component in multi-processor systems. 
Our attacks require that at least one memory module is shared between the attacker and the victim, which is the case even in the most restrictive settings. In these settings, attacker and victim cannot access the same memory cells, \ie we do not circumvent system-level memory isolation.
We do not make any assumptions on the cache, nor on the location of executing cores, nor on the availability of shared memory such as cross-VM memory deduplication.

First, we build a covert channel that achieves transmission rates of up to 2\,Mbps, which is three to four orders of magnitude faster than previously presented memory-bus based channels.
Second, we build a side channel that allows to automatically locate and monitor memory accesses, \eg user input or server requests, by performing template attacks.
Third, we show how the reverse-engineered mapping can be used to improve existing attacks. Existing Flush+Reload cache attacks use an incorrect cache-miss threshold, introducing noise and reducing the spatial accuracy. Knowledge of the DRAM address mapping also enables practical Rowhammer attacks on DDR4.

\subheading{Outline.}
The remainder of the paper is organized as follows. 
In Section~\ref{sec:background}, we provide background information on side channels on shared hardware, on DRAM, and on the Rowhammer attack. 
In Section~\ref{sec:definitions}, we provide definitions that we use throughout the paper.
In Section~\ref{sec:reveng}, we describe our two approaches to reverse engineer the DRAM addressing and we provide the reverse-engineered functions. 
In Section~\ref{sec:covert}, we build a high-speed cross-CPU DRAMA covert channel. 
In Section~\ref{sec:sidechannel}, we build a highly accurate cross-CPU DRAMA side channel attack.
In Section~\ref{sec:improving}, we show how the knowledge of the DRAM addressing improves cache attacks like Flush+Reload and we show how it makes Rowhammer attacks practical on DDR4 and more efficient on DDR3.
We discuss countermeasures against our attack in Section~\ref{sec:countermeasures}.
We conclude in Section~\ref{sec:conclusions}.

\section{Background and related work}\label{sec:background}
In this section, we  discuss existing covert and side channels and give an introduction to DRAM. Furthermore, we briefly explain the Rowhammer bug and its implications.

\subsection{Hardware covert and side channels} 
Attacks exploiting hardware sharing can be grouped into two categories. In side-channel attacks, an attacker spies on a victim and extracts sensitive information such as cryptographic keys.
In covert channels however, sender and receiver are actively cooperating to exchange information in a setting where they are not allowed to, \eg across isolation boundaries.

\subheading{Cache attacks.}
Covert and side channels using the CPU cache exploit the fact that cache hits are faster than cache misses. 
The methods Prime+Probe~\cite{Percival2005,Maurice2015DIMVA,Liu2015SP} and Flush+Reload~\cite{Yarom2014Usenix,Irazoqui2014RAID,Benger2014} have been presented to either build covert or side channels. These two methods work at a different granularity: Prime+Probe can spy on cache sets, while Flush+Reload has the finer granularity of a cache line but requires shared memory, such as shared libraries or memory deduplication.

Attacks targeting the last-level cache are cross-core, but require the sender and receiver to run on the same physical CPU. 
Gruss~\etal\cite{Gruss2016Flush} implemented cross-core covert channels using Prime+Probe and Flush+Reload as well as a new one, Flush+Flush, with the same protocol to normalize the results. The covert channel using Prime+Probe achieves 536\,Kbps, Flush+Reload 2.3\,Mbps, and Flush+Flush 3.8\,Mbps. The most recent cache attack by Irazoqui~\etal\cite{Irazoqui2015cpca} exploits cache coherency mechanisms and work across processors. It however requires shared memory.

An undocumented function maps physical addresses to the slices of the last-level cache. However, this function has been reverse engineered in previous work~\cite{Maurice2015RAID,Inci2015iacr898,Yarom2015iacr905}, enhancing existing attacks and enabling attacks in new environments.

\subheading{Memory and memory bus.}
Xiao~\etal\cite{Xiao2013} presented a covert channel that exploits memory deduplication. In order to save memory, the hypervisor searches for identical pages in physical memory and merges them across VMs to a single read-only physical page. Writing to this page triggers a copy-on-write page fault, incurring a significantly higher latency than a regular write access.
The authors built a covert channel that achieves up to 90\,bps, and 40\,bps on a system under memory pressure.
Wu~\etal\cite{Wu2014} proposed a bus-contention-based covert channel, that uses
atomic memory operations locking the memory bus. This covert channel achieves a raw bandwidth of 38\,Kbps between two VMs, with an effective capacity of 747\,bps with error correction.

\subsection{DRAM organization}\label{subsec:dram_org}
Modern DRAM is organized in a hierarchy of channels, DIMMs, ranks, and banks.
A system can have one or more \textit{channels}, which are physical links between the DRAM modules and the memory controller. Channels are independent and can be accessed in parallel. This allows distribution of the memory traffic, increasing the bandwidth, and reducing the latency in many cases.
Multiple \textit{Dual Inline Memory Modules} (\textit{DIMMs}), which are the physical memory modules attached to the mainboard, can be connected to each channel. A DIMM typically has one or two \textit{ranks}, which often correspond to the front and back of the physical module.
Each rank is composed of \textit{banks}, typically 8 on DDR3 DRAM and 16 on DDR4 DRAM. In the case of DDR4, banks are additionally grouped into \textit{bank groups}, \eg 4 bank groups with 4 banks each.
Banks finally contain the actual memory arrays which are organized in \textit{rows} (typically $2^{14}$ to $2^{17}$) and \textit{columns} (often $2^{10}$). On PCs, the DRAM word size and bus width is 64 bits, resulting in a typical row size of 8\,KB.
As channel, rank and bank form a hierarchy, two addresses can only be physically adjacent in the DRAM chip if they are in the same channel, DIMM, rank and bank. In this case we just use the term same bank.

The memory controller, which is integrated into modern processors, translates physical addresses to channels, DIMMs, ranks, and banks. AMD publicly documents the addressing function used by its products (see, \eg \cite[p. 345]{AMD-manual}), however to the best of our knowledge Intel does not.
The mapping for one Intel Sandy Bridge machine in one memory configuration has been reverse engineered by Seaborn~\cite{Seaborn2015DRAMmap}. However, Intel has changed the mapping used in its more recent microarchitectures. Also, the mapping necessarily differs when using other memory configurations, \eg a different number of DIMMs.

\subheading{The row buffer.}
Apart from the memory array, each bank also features a row buffer between the DRAM cells and the memory bus. From a high-level perspective, it behaves like a directly-mapped cache and stores an entire DRAM row.
Requests to addresses in the currently active row are served directly from this buffer.
If a different row needs to be accessed, then the currently active row is first closed (with a pre-charge command) and then the new row is fetched (with a row-activate command).
We call such an event a row conflict.
Naturally, such a conflict leads to significantly higher access times compared to requests to the active row. This timing difference will later serve as the basis for our attacks and for the software-based reverse-engineering method.
Note that after each refresh operation, a bank is already in the pre-charged state. In this case, no row is currently activated.

Independently of our work, Hassan~\etal~\cite{Hassan2015} also proposed algorithms to reverse engineer DRAM functions based on timing differences. However, their approach requires customized hardware performance-monitoring units. Thus, they tested their approach only in a simulated environment and not on real systems. Concurrently to our work, Xiao~\etal\cite{Xiao2016} proposed a method to reverse engineer DRAM functions based on the timing differences caused by row conflicts. Although their method is similar to ours, their focus is different, as they used the functions to then perform Rowhammer attacks across VMs.

\subheading{DRAM organization for multi-CPU systems.}
In modern multi-CPU server systems, each CPU features a dedicated memory controller and attached memory.
The DRAM is still organized in one single address space and is accessible by all processors. Requests for memory attached to other CPUs are sent over the CPU interconnect, \eg Intel's QuickPath Interconnect (QPI). This memory design is called Non-Uniform Memory Access (NUMA), as the access time depends on the memory location.

On our dual Haswell-EP setup, the organization of this single address space can be configured for the expected workload. In \emph{interleaved mode}, the memory is split into small slices which are spliced together in an alternating fashion.
In \emph{non-interleaved} mode, each CPUs memory is kept in one contiguous physical-address block. For instance, the lower half of the address space is mapped to the first CPUs memory, whereas the upper half is mapped to the second CPUs memory.

\subsection{The Rowhammer bug}
The increasing DRAM density has led to physically smaller cells, which can thus store smaller charges.
As a result, the cells have a lower noise margin and the level of parasitic electrical interaction is potentially higher, resulting in the so-called Rowhammer bug~\cite{Kim2014ISCA,Huang2012,Park2014}.

This bug results in corruption of data, not in rows that are directly accessed, but rather in adjacent ones. When performing random memory accesses, the probability for such faults is virtually zero. However, it rises drastically when performing accesses in a certain pattern. Namely, flips can be caused by frequent activation (\emph{hammering}) of adjacent rows.
As data needs to be served from DRAM and not the cache, an attack needs to either flush data from the cache using the \texttt{clflush} instruction in native environments~\cite{Kim2014ISCA}, or using cache eviction in other more restrictive environments, \eg JavaScript~\cite{Gruss2016Row}.

Seaborn~\cite{Seaborn2015row} implemented two attacks that exploit the Rowhammer bug, showing the severity of faulting single bits for security. 
The first exploit is a kernel privilege escalation on a Linux system, caused by a bit flip in a page table entry.
The second one is an escape of Native Client sandbox caused by a bit flip in an instruction sequence for indirect jumps.

\section{Definitions}\label{sec:definitions}
In this section we provide definitions for the terms \emph{row hit} and \emph{row conflict}. These definitions provide the basis for our reverse engineering as well as the covert and side channel attacks.

Every physical memory location maps to one out of many rows in one out of several banks in the DRAM.
Considering a single access to a row $i$ in a bank there are two major possible cases: 
\begin{compactenum}
\item The row $i$ is already opened in the row buffer. We call this case a \emph{row hit}.
\item A different row $j \neq i$ in the same bank is opened. We call this case a \emph{row conflict}. 
\end{compactenum}

Considering frequent alternating accesses to two (or more) addresses we distinguish three cases:
\begin{compactenum}
\item The addresses map to different banks. In this case the accesses are independent and whether the addresses have the same row indices has no influence on the timing. Row hits are likely to occur for the accesses, \ie access times are low.
\item The addresses map to the same row $i$ in the same bank. The probability that the row stays open in between accesses is high, \ie access times are low.
\item The addresses map to the different rows $i \neq j$ in the same bank.
Each access to an address in row $i$ will close row $j$ and vice versa. Thus, row conflicts occur for the accesses, \ie access times are high.
\end{compactenum}

\begin{figure*}[tb]
  \centering
  \includegraphics{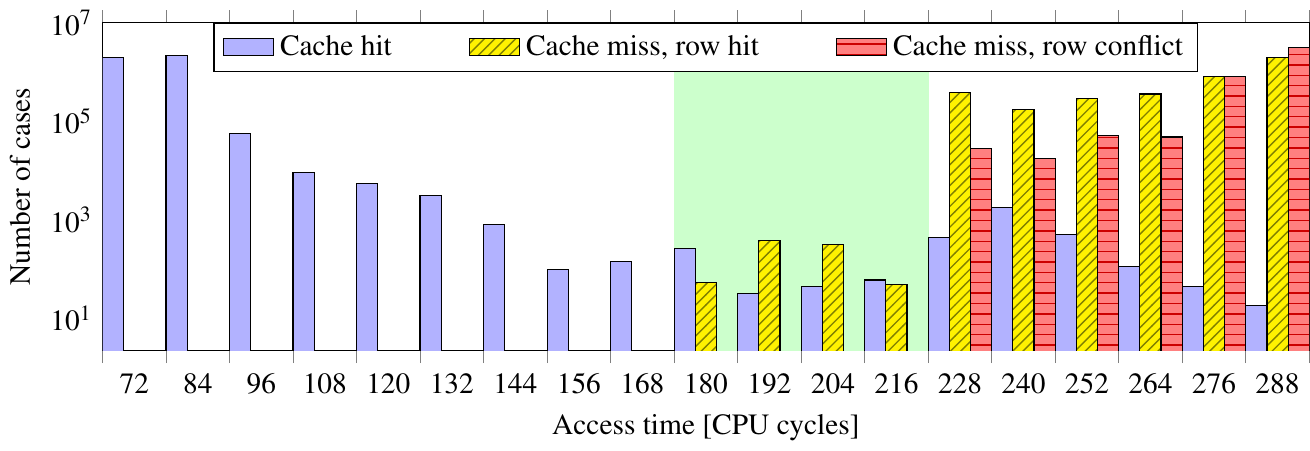}
  \caption{Histogram for cache hits and cache misses divided into row hits and row conflicts on the Ivy Bridge i5 test system. Measurements were performed after a short idle period to simulate non-overlapping accesses by victim and spy. From 180 to 216 cycles row hits occur, but no row conflicts.}
  \label{fig:hist_cache_comparison}
\end{figure*}

To measure the timing differences of row hits and row conflicts, data has to be flushed from the cache.
Figure~\ref{fig:hist_cache_comparison} shows a comparison of standard histograms of access times for cache hits and cache misses. Cache misses are further divided into row hits and row conflicts. For this purpose an unrelated address in the same row was accessed to cause a row hit and an unrelated address in the same bank but in a different row was accessed to cause a row conflict. We see that from 180 to 216 cycles row hits occur, but no row conflicts (cf.\ highlighted area in Figure~\ref{fig:hist_cache_comparison}). In the remainder, we build different attacks that are based on this timing difference between row hits and row conflicts.

\section{Reverse engineering DRAM addressing}\label{sec:reveng}
In this section, we present our reverse engineering of the DRAM address mapping. We discuss two approaches, the first one is based on physical probing, whereas the second one is entirely software-based and fully automated. Finally, we present the outcome of our analysis, \ie the reverse-engineered mapping functions.
In the remainder of this paper, we denote with $a$ a physical memory address. $a_i$ denotes the $i$-th bit of an address.

\subsection{Linearity of functions}
The DRAM addressing functions are reverse engineered in two phases. First, a measuring phase and second, a subsequent solving phase. Our solving approaches require that the addressing functions are linear, \ie they are XORs of physical-address bits.

In fact, Intel used such functions in earlier microarchitectures. For instance, Seaborn \cite{Seaborn2015DRAMmap} reports that on his Sandy Bridge setup the bank address is computed by XORing the bits $a_{14}..a_{16}$ with the lower bits of the row number ($a_{18}..a_{20}$) (cf.\ Figure~\ref{fig:mapping_sandy}). This is done in order to minimize the number of row conflicts during runtime.
Intel also uses linear functions for CPU-cache addressing. Maurice~\etal\cite{Maurice2015RAID} showed that the \emph{complex addressing} function, which is used to select cache slices, is an XOR of many physical-address bits.

As it turns out, linearity holds on all our tested configurations. However, there are setups in which it might be violated, such as triple-channel configurations. We did not test such systems and leave a reverse engineering to future work.

\subsection{Reverse engineering using physical probing}\label{subsec:method-physical}
\begin{figure}[tb]
\centering
\includegraphics[width=1.0\linewidth]{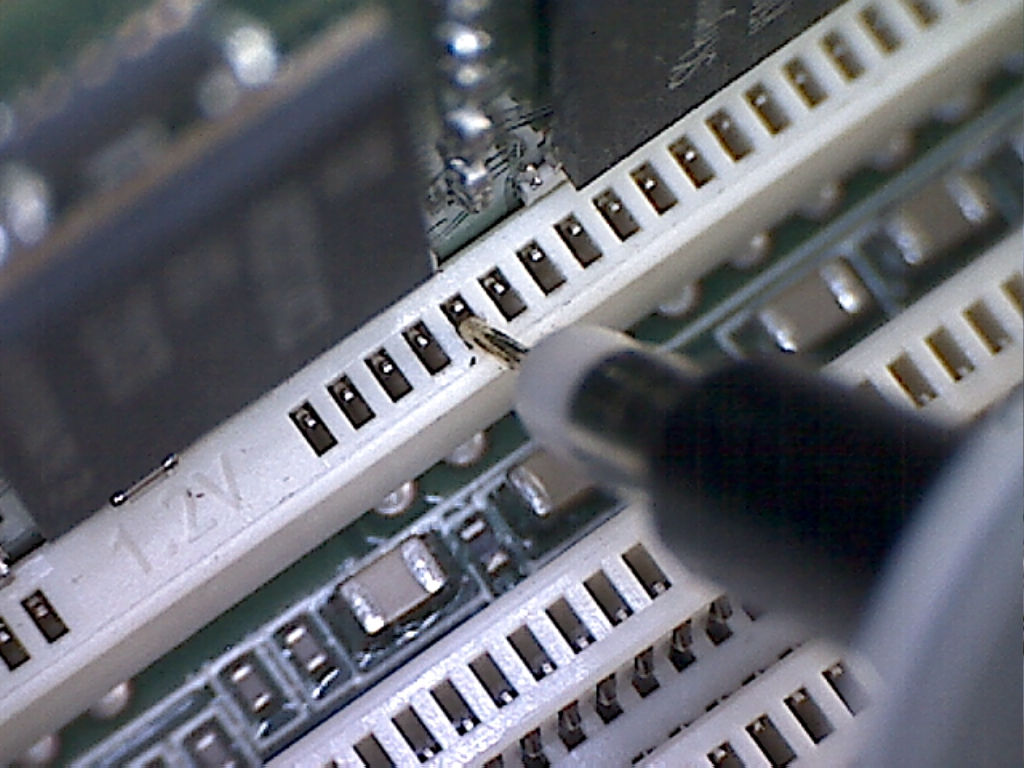}
\caption{Physical probing of the DIMM slot.}
\label{fig:phys_probing}
\end{figure}

Our first approach to reverse engineer the DRAM mapping is to physically probe the memory bus and to directly read the control signals.
As shown in Figure~\ref{fig:phys_probing}, we use a standard passive probe to establish contact with the pin at the DIMM slot.
We then repeatedly accessed a selected physical address\footnote{Resolving virtual to physical addresses requires root privileges in Linux. Given that we need physical access to the internals of the system, this is a very mild prerequisite.} and used a high-bandwidth oscilloscope to measure the voltage and subsequently deduce the logic value of the contacted pin. Note that due to the repeated access to a single address, neither a timely location of specific memory requests nor distinguishing accesses to the chosen address from other random ones is required.

We repeated this experiment for many selected addresses and for all pins of interest, namely the bank-address bits (BA0, BA1, BA2 for DDR3 and BG0, BG1, BA0, BA1 for DDR4) for one DIMM and the chip select CS for half the DIMMs.

For the solving phase we use the following approach. Starting from the top-layer (channel or CPU addressing) and drilling down, for each DRAM addressing function we create an over-defined system of linear equations in the physical address bits.
The left-hand-side of this system is made up of the relevant tested physical addresses. For instance, for determining the bank functions we only use addresses that map to the contacted DIMMs channel.
The right-hand-side of the system of equations are the previously measured logic values for the respective address and the searched-for function. The logic values for CPU and channel addressing are computed by simply ORing all respective values for the chip-select pins.  
We then solve this system using linear algebra. The solution is the corresponding DRAM addressing function.

Obviously, this reverse-engineering approach has some drawbacks. First, expensive measurement equipment is needed. Second, it requires physical access to the internals of the tested machine.
However, it has the big advantage that the address mapping can be reconstructed for each control signal individually and exactly. Thus, we can determine the exact individual functions for the bus pins. Furthermore, every platform only needs to be measured only once in order to learn the addressing functions. Thus, an attacker does not need physical access to the concrete attacked system if the measurements are performed on a similar machine.

\subsection{Fully automated reverse engineering}\label{sec:timingreveng}
For our second approach to reverse engineer the DRAM mapping we exploit the fact that row conflicts lead to higher memory access times.
We use the resulting timing differences to find sets of addresses that map to the same bank but to a different row. Subsequently, we determine the addressing functions based on these sets. The entire process is fully automated and runs in unprivileged and possibly restricted environments.

\begin{figure}[tb]
  \centering
\begin{tikzpicture}[scale=0.9]
\begin{axis}[
xlabel={Access time [CPU cycles]},
ylabel={Proportion of cases},
width=\hsize,
height=3.8cm,
ymin=0,
ybar,
bar width=3,
]
\addplot+[] table[x=x,y=y] {histogram_updated.csv};
\end{axis}
\end{tikzpicture}
  \caption{Histogram of average memory access times for random address pairs on our Haswell test system. A clear gap separates the majority of address pairs causing no row conflict (lower access times), because they map to different banks, from the few address pairs causing a row conflict (higher access times), because they map to different rows in the same bank.
  }
  \label{fig:histograms}
\end{figure}
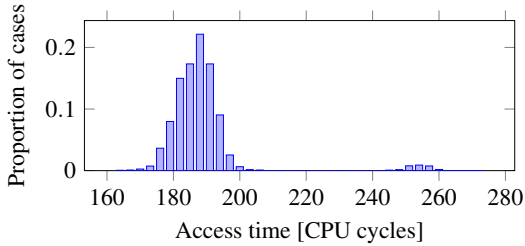

\subheading{Timing analysis.}
In the first step, we aim to find same-bank addresses in a large array mapped into the attackers' address space.
For this purpose, we perform repeated alternating access to two addresses and measure the average access time. We use \texttt{clflush} to ensure that each access is served from DRAM and not from the CPU cache.
As shown in Figure~\ref{fig:histograms}, for some address pairs the access time is significantly higher than for most others. These pairs belong to the same bank but to different rows. The alternating access causes frequent row conflicts and consequently the high latency.

The tested pairs are drawn from an address pool, which is built by selecting random addresses from a large array. A small subset of addresses in this pool is tested against all others in the pool. The addresses are subsequently grouped into sets having the same channel, DIMM, rank, and bank.
We try to identify as many such sets as possible in order to reconstruct the addressing functions.

\subheading{Function reconstruction.}
In the second phase, we use the identified address sets to reconstruct the addressing functions. This reconstruction requires (at least partial) resolution of the tested virtual addresses to physical ones. Similar as later in Section\,\ref{subsec:covert-basic}, one can use either the availability of 2\,MB pages, 1\,GB pages, or privileged information such as the virtual-to-physical address translation that can be obtained through \verb|/proc/pid/pagemap| in Linux systems.

In the case of 2\,MB pages we can recover all partial functions up to bit $a_{20}$, as the lowest 21 bit of virtual and physical address are identical. On many systems the DRAM addressing functions do not use bits above $a_{20}$ or only few of them, providing sufficient information to mount covert and side-channel attacks later on. In the case of 1\,GB pages we can recover all partial functions up to bit $a_{30}$. This is sufficient to recover the full DRAM addressing functions on all our test systems. If we have full access to physical address information we will still ignore bits $a_{30}$ and upwards. These bits are typically only used for DRAM row addressing and they are very unlikely to play any role in bank addressing. Additionally, we ignore bits $(a_0..a_5)$ as they are used for addressing within a cache line.

The search space is then small enough to perform a brute-force search of linear functions within seconds. For this, we generate all linear functions that use exactly $n$ bits as coefficients and then apply them to all addresses in one randomly selected set. We start with $n=1$ and increment $n$ subsequently to find all functions.
Only if the function has the same result for all addresses in a set, we test this potential function on all other sets. However, in this case we only pick one address per set and test whether the function is constant over all sets. If so, the function is discarded.
We obtain a list of possible addressing functions that also contains linear combinations of the actual DRAM addressing functions. We prioritize functions with a lower number of coefficients, \ie we remove higher-order functions which are linear combinations of lower-order ones.
Depending on the random address selection, we now have a complete set of correct addressing functions. We verify the correctness either by comparing it to the results from the physical probing, or by performing a software-based test, \ie verifying the timing differences on a larger set of addresses, or verifying that usage of the addressing functions in Rowhammer tests increases the number of bit flips per second by a factor that is the number of sets we found.

Compared to the probing approach, this purely software-based method has significant advantages. It does not require any additional measurement equipment and can be executed on a remote system. We can identify the functions even from within VMs or sandboxed processes if 2\,MB or 1\,GB pages are available. Furthermore, even with only 4\,KB pages we can group addresses into sets that can be directly used for covert or side channel attacks.
This software-based approach also allows reverse engineering in settings where probing is not easily possible anymore, such as on mobile devices with hard-wired ball-grid packages. Thus, it allowed us to reverse engineer the mapping on current ARM processors.

One downside of the software-based approach is that it cannot recover the exact labels (BG0, BA0, ...) of the functions. Thus, we can only guess whether the reconstructed function computes a bank address bit, rank bit, or channel bit. Note that assigning the correct labels to functions is not required for any of our attacks.

\subsection{Results}\label{subsec:results}
We now present the reverse-engineered mappings for all our experimental setups. We analyzed a variety of systems (Table~\ref{tbl:machines}), including a dual-CPU Xeon system, that can often be found in cloud systems, and multiple current smartphones. Where possible, we used both presented reverse-engineering methods and cross-validated the results.

\begin{table}[tb]
  \begin{center}
    \caption{Experimental setups.}\label{tbl:machines}
    {    \small
    \footnotesize
      \begin{tabular}{ccc}
        \toprule
        CPU / SoC & Microarch. &  Mem.\\
        \midrule
        i5-2540M & Sandy Bridge & DDR3 \\
        i5-3230M & Ivy Bridge & DDR3 \\
        i7-3630QM & Ivy Bridge & DDR3 \\
        i7-4790 & Haswell & DDR3 \\
        i7-6700K & Skylake & DDR4 \\
        2x Xeon E5-2630 v3 & Haswell-EP & DDR4 \\
        Qualcomm Snapdragon S4 Pro & ARMv7 & LPDDR2 \\
                Samsung Exynos 5 Dual & ARMv7 & LDDDR3 \\
                Qualcomm Snapdragon 800 & ARMv7 & LPDDR3 \\
                Qualcomm Snapdragon 820 & ARMv8-A & LPDDR3 \\
                Samsung Exynos 7420 & ARMv8-A & LPDDR4 \\
        \bottomrule
      \end{tabular}
    }
  \end{center}
\end{table}

We found that the basic scheme is always as follows. On PCs, the memory bus is 64 bits wide, yet the smallest addressable unit is a byte. Thus, the three lower bits $(a_0..a_2)$ of the physical address are used as byte index into a 64-bit (8-byte) memory word and they are never transmitted on the memory bus. Then, the next bits are used for column selection. One bit in between is used for channel addressing. The following bits are responsible for bank, rank, and DIMM addressing. The remaining upper bits are used for row selection.

\begin{figure}[h!]
  \begin{subfigure}[b]{\columnwidth}
    \centering
    \includegraphics[scale=1.1]{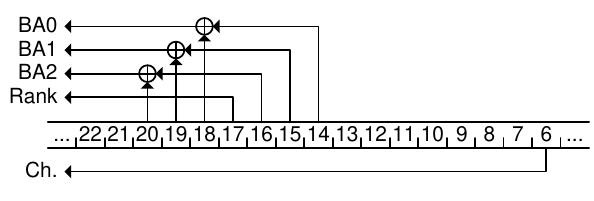}
    \caption{Sandy Bridge -- DDR3 \cite{Seaborn2015DRAMmap}.}
    \label{fig:mapping_sandy}
  \end{subfigure}
  ~
  
  \begin{subfigure}[b]{\columnwidth}
    \centering
    \includegraphics[scale=1.1]{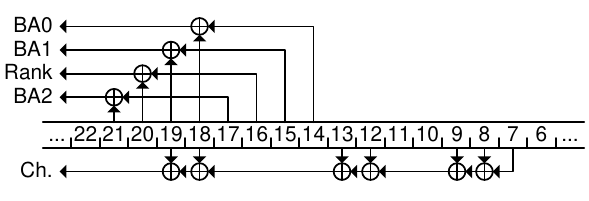}
    \caption{Ivy Bridge / Haswell -- DDR3.}
    \label{fig:mapping_haswell}
  \end{subfigure}
  ~
  
  \begin{subfigure}[b]{\columnwidth}
    \centering
    \includegraphics[scale=1.1]{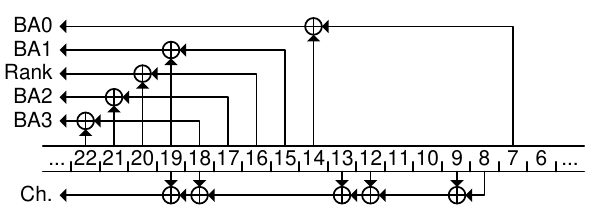}
    \caption{Skylake -- DDR4.}
    \label{fig:mapping_skylake}
  \end{subfigure}
  
  \begin{subfigure}[b]{\columnwidth}
    \centering
    \includegraphics[scale=1.1]{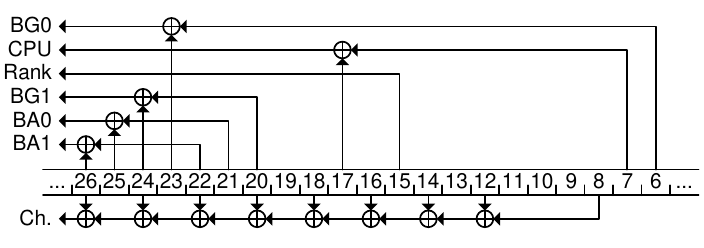}
    \caption{Dual Haswell-EP (Interleaved Mode) -- DDR4.}
    \label{fig:mapping_haswell_ep}
  \end{subfigure}
  
  \begin{subfigure}[b]{\columnwidth}
    \centering
    \includegraphics[scale=1.1]{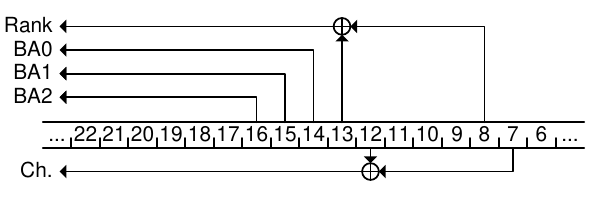}
    \caption{Samsung Exynos 7420 -- LPDDR4.}
    \label{fig:mapping_armv8}
  \end{subfigure}
  
  \caption{Reverse engineered dual channel mapping (1 DIMM per channel) for different architectures.}
  \label{fig:mappings_examples}
\end{figure}

The detailed mapping, however, differs for each setup. To give a quick overview of the main differences, we show the mapping of one selected memory configuration for multiple Intel microarchitectures and ARM-based SoCs in Figure~\ref{fig:mappings_examples}. Here we chose a configuration with two equally sized DIMMs in dual-channel configuration, as it is found in many off-the-shelf consumer PCs. All our setups use dual-rank DIMMs and use 10 bits for column addressing.
Figure~\ref{fig:mapping_sandy} shows the mapping on the Sandy Bridge platform, as reported by Seaborn~\cite{Seaborn2015DRAMmap}. Here, only $a_6$ is used to select the memory channel, $a_{17}$ is used for rank selection. The bank-address bits are computed by XORing bits $a_{14}..a_{16}$ with the lower bits of the row index ($a_{18}..a_{20}$).

The channel selection function changed with later microarchitectures, such as Ivy Bridge and Haswell. As shown in Figure~\ref{fig:mapping_haswell}, the channel-selection bit is now computed by XORing seven bits of the physical address. Further analysis showed that bit $a_7$ is used exclusively, \ie it is not used as part of the row- or column address. Additionally, rank selection is now similar to bank addressing and also uses XORs.

Our Skylake test system uses DDR4 instead of DDR3. Due to DDR4's introduction of bank grouping and the doubling of the available banks (now 16), the addressing function necessarily changed again. As shown in Figure~\ref{fig:mapping_skylake}, $a_7$ is not used for channel selection anymore, but for bank addressing instead.

Figure~\ref{fig:mapping_haswell_ep} depicts the memory mapping of a dual-CPU Haswell-EP system equipped with DDR4 memory. It uses 2 modules in dual-channel configuration \emph{per CPU} (4 DIMMs in total).
In interleaved mode (cf.\ Section~\ref{subsec:dram_org}), the chosen CPU is determined as $a_7 \oplus a_{17}$. Apart from the different channel function, there is also a difference in the bank addressing, \ie bank addressing bits are shifted. The range of bits used for row indexing is now split into address bits $(a_{17}..a_{19})$ and $a_{23}$ upwards.

\begin{table*}[th]
{    \centering
      
  \caption{Reverse engineered DRAM mapping on all platforms and configurations we analyzed via physical probing or via software analysis.
These tables list the bits of the physical address that are XORed. For instance, for the entry (13, 17) we have $a_{13} \oplus a_{17}$.}
  \label{tab:mapping}
  
  \begin{subtable}[b]{\hsize}
    \centering
    \small
    \footnotesize
    \caption{DDR3}
    \label{tab:mapping3}
    \begin{tabular}{cc c !{\vrule}  ccccccc}
    \toprule
      CPU & Ch. & DIMM/Ch. & BA0 & BA1 & BA2 & Rank & DIMM & Channel\\
\midrule
      Sandy Bridge & 1 & 1 & 13, 17 &14, 18 & 15, 19 & 16 & - & -\\
      Sandy Bridge~\cite{Seaborn2015DRAMmap} & 2 & 1 & 14, 18 & 15, 19 & 16, 20 & 17 & - & 6\\
      \hdashline[0.5pt/3pt]
      \multirow{4}{*}{Ivy Bridge/Haswell} & 1 & 1 & 13, 17 & 14, 18 & 16, 20 & 15, 19& - & - \\
      & 1 & 2 & 13, 18 & 14, 19 & 17, 21 & 16, 20 & 15 & - \\
      & 2 & 1 & 14, 18 & 15, 19 & 17, 21 & 16, 20 & - & 7, 8, 9, 12, 13, 18, 19 \\
      & 2 & 2 & 14, 19 & 15, 20 & 18, 22 & 17, 21 & 16 & 7, 8, 9, 12, 13, 18, 19 \\
      \bottomrule
    \end{tabular}
  \end{subtable}
  
  \bigskip
  
  \begin{subtable}[b]{\linewidth}
    \centering
      \small
      \footnotesize
      \caption{DDR4}
      \label{tab:mapping4}
      \begin{tabular}{cc c !{\vrule} ccccccc}
\toprule
        CPU & Ch. & DIMM/Ch. & BG0 & BG1 & BA0 & BA1 & Rank & CPU & Channel\\
\midrule
        Skylake$^{\dagger}$  & 2 & 1 & 7, 14 & 15, 19 & 17, 21 & 18, 22 & 16, 20 &-& 8, 9, 12, 13, 18, 19\\ \hdashline[0.5pt/3pt]
        2x Haswell-EP & 1 & 1 & 6, 22 & 19, 23 & 20, 24 & 21, 25 & 14 & 7, 17 & -\\
        (interleaved) & 2 & 1 & 6, 23 & 20, 24 & 21, 25 & 22, 26 & 15 & 7, 17&  8, 12, 14, 16, 18, 20, 22, 24, 26 \\ \hdashline[0.5pt/3pt]        
        2x Haswell-EP & 1 & 1 & 6, 21 & 18, 22 & 19, 23 & 20, 24 & 13 & - & -\\
        (non-interleaved) & 2 & 1 & 6, 22 & 19, 23, & 20, 24 & 21, 25 & 14 & - &  7, 12, 14, 16, 18, 20, 22, 24, 26 \\ \bottomrule
      \end{tabular}
  \end{subtable}
  \smallskip

        \begin{subtable}[b]{\linewidth}
                \centering
                        \small
                        \footnotesize
                        \caption{LPDDR2,3,4}
                        \label{tab:mapping_arm}
                        \begin{tabular}{cc!{\vrule} cccccc}
\toprule
                                CPU & Ch. & BA0 & BA1 & BA2 & Rank & Channel\\
\midrule
                                Qualcomm Snapdragon S4 Pro$^{\dagger}$ & 1 & 13 & 14 & 15 & 10 & - \\
                                Samsung Exynos 5 Dual$^{\dagger}$ & 1 & 13 & 14 & 15 & 7 & - \\
                                Qualcomm Snapdragon 800/820$^{\dagger}$ & 1 & 13 & 14 & 15 & 10 & - \\
                                Samsung Exynos 7420$^{\dagger}$  & 2 & 14 & 15 & 16 & 8, 13 & 7, 12\\
\bottomrule
                        \end{tabular}
        
        \end{subtable}
        \smallskip
}

\vspace{0.1cm}
\begin{center}
\footnotesize{
$^{\dagger}$ Software analysis only. Labeling of functions is based on results of other platforms.
}
\end{center}
\end{table*}

The mapping used on one of our mobile platforms, a Samsung Galaxy S6 with an Exynos 7420 ARMv8-A SoC and LPDDR4 memory, is much simpler (cf.\ Figure~\ref{fig:mapping_armv8}). Here physical address bits are mapped directly to bank address bits. Rank and channel are computed with XORs of only two bits each. The bus width of LPDDR4 is 32 bits, so only the two lowest bits are used for byte indexing in a memory word.

Table~\ref{tab:mapping} shows a comprehensive overview of all platforms and memory configurations we analyzed. As all found functions are linear, we simply list the index of the physical address bits that are XORed together.
With the example of the Haswell microarchitecture, one can clearly see that the indices are shifted to accommodate for the different memory setups. For instance, in single-channel configurations $a_7$ is used for column instead of channel selection, which is why bank addressing starts with $a_{13}$ instead of $a_{14}$.

\section{A high-speed cross-CPU covert channel}\label{sec:covert}
In this section, we present a first DRAMA attack, namely a high-speed cross-CPU covert channel that does not require shared memory. Our channel exploits the row buffer, which behaves like a directly-mapped cache.
Unlike cache attacks, the only prerequisite is that two communicating processes have access to the same memory module.

\subsection{Basic concept}\label{subsec:covert-basic}
\begin{figure}[tb]
\centering
\includegraphics[width=1.0\linewidth]{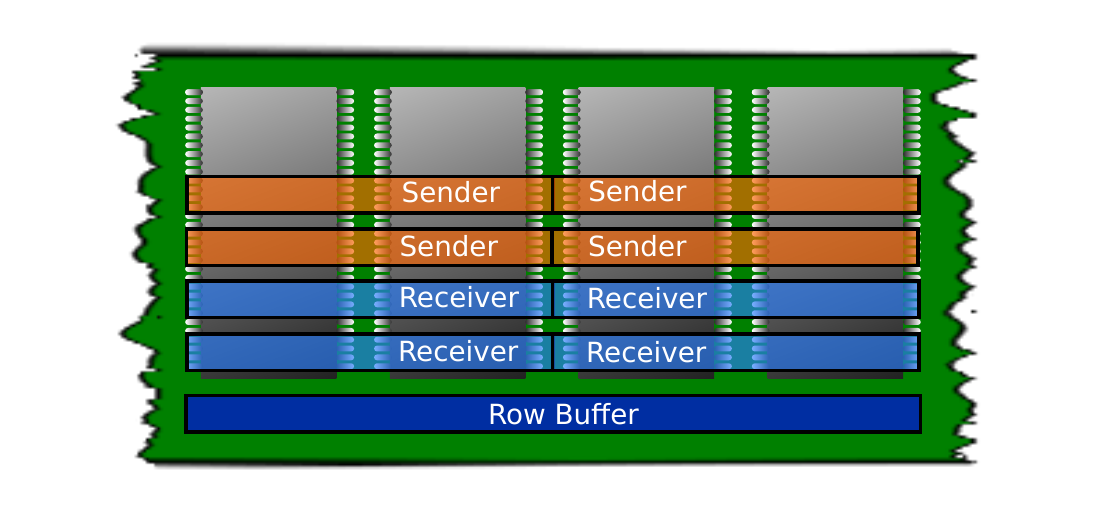}
\caption{The sender occupies rows in a bank to trigger row conflicts. The receiver occupies rows in the same bank to observe these row conflicts.}
\label{fig:row_miss_attack}
\end{figure}

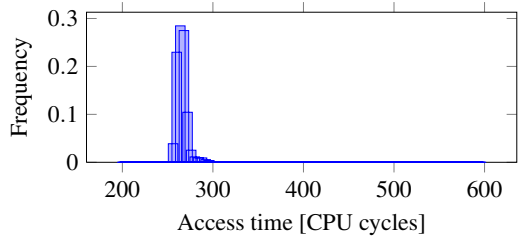
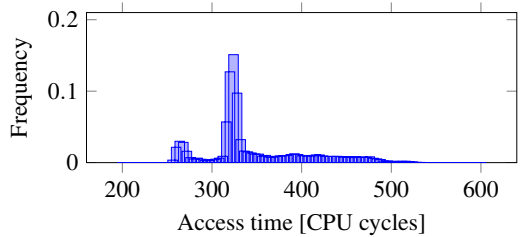
\begin{figure}[tb]
  \centering
    \begin{subfigure}[b]{\columnwidth}
      \begin{tikzpicture}[scale=0.9]
      \begin{axis}[
      line width=0.2pt,
      xlabel={Access time [CPU cycles]},
      ylabel={Frequency},
      width=\hsize,
      height=3.8cm,
      ymin=0,
      ybar,
      bar width=4,
      ]
      \addplot+[] table[x=x,y=y] {deltabuf0.csv};
      \end{axis}
      \end{tikzpicture}
      \caption{Sender inactive on bank: sending a 0.}
      \label{fig:covert_channel_send0}
    \end{subfigure}
    
    \medskip
    
    \begin{subfigure}[b]{\columnwidth}
      \begin{tikzpicture}[scale=0.9]
      \begin{axis}[
      line width=0.2pt,
      xlabel={Access time [CPU cycles]},
      ylabel={Frequency},
      width=\hsize,
      height=3.8cm,
      ymin=0,
      ymax=0.21,
      ytick={0, 0.1, 0.2},
      ybar,
      bar width=4,
      ]
      \addplot+[] table[x=x,y=y] {deltabuf1.csv};
      \end{axis}
      \end{tikzpicture}
      \caption{Sender active on bank: sending a 1.}
      \label{fig:covert_channel_send1}
    \end{subfigure}
    
  \caption{Timing differences between active and non-active sender (on one bank), measured on the Haswell i7 test system.}
  \label{fig:timing-diff}
\end{figure}

\begin{figure}[tb]
  \centering
  \begin{tikzpicture}
  \pgfplotsset{every axis legend/.append style={at={(0.5,1.3)},anchor=north}}
  \begin{axis}[
  legend columns=4,
  xlabel={Time [$\mu$s]},
  ylabel=Access time,
  ytick pos=left,
  scaled y ticks = false,
  width=\hsize,
  height=3.8cm,
  ]
  \addplot+[no marks, color=blue] table[x=x,y=y] {raw.csv};
  \end{axis}
  \end{tikzpicture}
  \caption{Covert channel transmission on one bank, cross-CPU and cross-VM on a Haswell-EP server. The time frame for one bit is 50$\mu$s.}
  \label{fig:transmission_vm}
\end{figure}
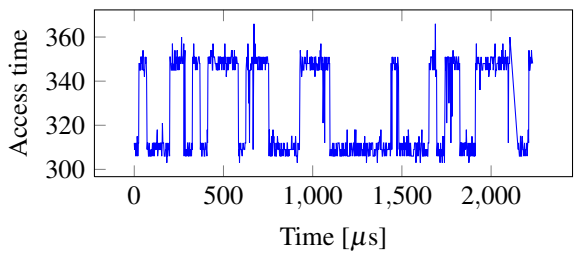

Our covert channel exploits timing differences caused by row conflicts. Sender and receiver occupy different rows in the same bank as illustrated in Figure~\ref{fig:row_miss_attack}. The receiver process continuously accesses a chosen physical address in the DRAM and measures the average access time over a few accesses.
If the sender process now continuously accesses a different address in the same bank but in a different row, a row conflict occurs.
This leads to higher average access times in the receiver process. Bits can be transmitted by switching the activity of the sender process in the targeted bank on and off. This timing difference is illustrated in Figure~\ref{fig:timing-diff}, an exemplary transmission is shown in Figure~\ref{fig:transmission_vm}.
The receiver process distinguishes the two values based on the mean access time. We assign a logic value of 0 to low access times (the sender is inactive) and a value of 1 to high access times (the sender is active).

Each (CPU, channel, DIMM, rank, bank) tuple can be used as a separate transmission channel.
However, a high number of parallel channels leads to increased noise. Also, there is a strict limit on the usable bank parallelism. Thus, optimal performance is achieved when using only a subset of available tuples.
Transmission channels are unidirectional, but the direction can be chosen for each one independently. Thus, two-way communication is possible.

To evaluate the performance of this new covert channel, we created a proof-of-concept implementation. We restrict ourselves to unidirectional communication, \ie there is one dedicated sender and one dedicated receiver.

The memory access time is measured using \texttt{rdtsc}. The memory accesses are performed using \texttt{volatile} pointers. In order to cause a DRAM access for each request, data has to be flushed from the cache using \texttt{clflush}.

\subheading{Determining channel, rank, and bank address.}
In an agreement phase, all parties need to agree on the set of (channel, DIMM, rank, bank) tuples that are used for communication. This set needs to be chosen only once, all subsequent communication can use the same set.
Next, both sender and receiver need to find at least one address in their respective address space for each tuple. Note that some operating systems allow unprivileged resolution of virtual to physical addresses. In this case, finding correct addresses is trivial.

However, on Linux, which we used on our testing setup, unprivileged address resolution is not possible. Thus, we use the following approach. 
As observed in previous work~\cite{Gruss2015esorics,Gruss2016Row}, system libraries and the operating system assign 2\,MB pages for arrays which are significantly larger than 2\,MB. 
On these pages, the 21 lowest bits of the virtual address and the physical address are identical. Depending on the hardware setup, these bits can already be sufficient to fully determine bank, rank and channel address.
For this purpose, both processes request a large array. The start of this array is not necessarily aligned with a 2\,MB border. Memory before such a border is allocated using 4\,KB pages. We skip to the next 2\,MB page border by choosing the next virtual address having the 21 lowest bits set to zero.

On systems that also use higher bits, an attacker can use the following approach, which we explain on the example of the mapping shown in Figure~\ref{fig:mapping_haswell}. There an attacker cannot determine the BA2 bit by just using 2\,MB pages. 
Thus, the receiving process selects addresses with chosen BA0, BA1, rank, and channel, but unknown BA2 bit. The sender now accesses addresses for both possibilities of BA2, \eg by toggling $a_{17}$ between consecutive reads. Thus, only each second access in the sending process targets the correct bank. Yet, due to bank parallelism this does not cause a notable performance decrease.
Note however that this approach might not work if the number of unknown bank-address bits is too high.

In a virtualized environment, even a privileged attacker is able to retrieve only the \emph{guest physical address}, which is further translated into the real physical address by the memory management unit. However, if the host system uses 1\,GB pages for the second-level address translation (to improve efficiency), then the lowest 30 bits of the guest physical address are identical to the real physical address. Knowledge of these bits is sufficient on all systems we analyzed to use the full DRAM addressing functions.

Finally, the covert channel could also be built without actually reconstructing the DRAM addressing functions. Instead of determining the exact bank address, it can rely solely on the same-bank sets retrieved in Section~\ref{sec:timingreveng}. In an initialization phase, both sender and receiver perform the timing analysis and use it to build sets of same-bank addresses. Subsequently, the communicating parties need to synchronize their sets, \ie they need to agree on which of them is used for transmission. This is done by sending predefined patterns over the channel. After that, the channel is ready for transmission. Thus, it can be established without having any information on the mapping function nor on the physical addresses.

\subheading{Synchronization.}
In our proof-of-concept implementation, one set of bits (a data block) is transmitted for a fixed time span which is agreed upon before starting communication. Decreasing this period increases the raw bitrate, but it also increases the error rate, as shown in Figure~\ref{fig:bitrate}.

For synchronizing the start of these blocks we employ two different mechanisms.
If sender and receiver run natively, we use the wall clock as means of synchronization. Here blocks start at fixed points in time. If, however, sender and receiver run in two different VMs, then a common (or perfectly synchronized) wall clock is typically not available. In this case, the sender uses one of the transmission channels to transmit a clock signal which toggles at the beginning of each block. The receiver then recovers this clock and can thus synchronize with the sender.

We employ multiple threads for both the sender and receiver processes to achieve optimal usage of the memory bus. Thus, memory accesses are performed in parallel, increasing the performance of the covert channel.

\subsection{Evaluation}
We evaluated the performance of our covert-channel implementation on two systems.
First, we performed tests on a standard desktop PC featuring an Intel i7-4790 CPU with Haswell microarchitecture. It was equipped with 2 Kingston DDR3 KVR16N11/8 dual-rank 8\,GB DIMMs in dual-channel configuration. 
The system was mostly idle during the tests, \ie there were no other tasks causing significant load on the system. The DRAM clock was set to its default of 800\,MHz (DDR3-1600).

Furthermore, we also tested the capability of cross-CPU transmission on a server system. Our setup has two Intel Xeon E5-2630 v3 (Haswell-EP microarchitecture). It was equipped with a total of 4 Samsung M393A2G40DB0-CPB DDR4 registered ECC DIMMs. Each CPU was connected to two DIMMs in dual-channel configuration and NUMA was set to interleaved mode.
The DRAM frequency was set to its maximum supported value (DDR4-1866).

For both systems, we evaluated the performance in both a native scenario, \ie both processes run natively, and in a cross-VM scenario. We transmit 8 bits per block (use 8 (CPU, channel, DIMM, rank, bank) tuples) in the covert channel and run 2 threads in both the sender and the receiver process. Every thread is scheduled to run on different CPU cores, and in the case of the Xeon system, sender and receiver run on different physical CPUs.

We tested our implementation with a large range of measurement intervals. For each one, we measure the raw channel capacity and the bit error probability. While the raw channel capacity increases proportionally to the reduction of the measurement time, the bit error rate increases significantly if the measurement time is too short.
In order to find the best transmission rate, we use the channel capacity as metric. When using the binary symmetric channel model, this metric is computed by multiplying the raw bitrate with $1- H(e)$, with $e$ the bit error probability and $H(e) = -e\cdot\log_2(e) - (1-e)\cdot \log_2(1-e)$ the binary entropy function.

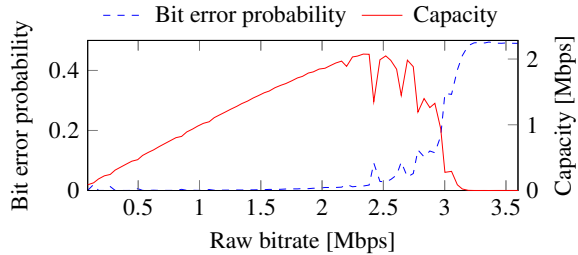
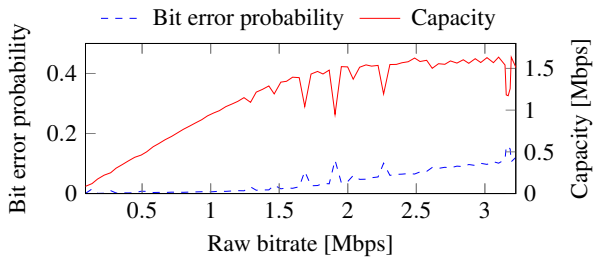
\begin{figure}[tb] 
  \begin{subfigure}[b]{\columnwidth}
    \centering
\begin{tikzpicture}[scale=0.9]
\pgfplotsset{every axis legend/.append style={at={(0.5,1.3)},anchor=north,draw=none}}
\begin{axis}[
legend columns=2,
xlabel={Raw bitrate [Mbps]},
ylabel=Bit error probability,
ytick pos=left,
xmin=0.088,
xmax=3.598,
ymin=0,
ymax=0.5,
scaled y ticks = false,
width=\hsize,
height=3.8cm,
]
\addplot+[no marks, color=blue, dashed] table[x=mbits,y=ber] {covert_channel_pc.csv};
\addlegendentry{~Bit error probability \,}
\addplot+[no marks, color=red, draw=none] coordinates {(-5,-5)};
\addlegendentry{~Capacity}
\end{axis}
\begin{axis}[
legend columns=2,
xlabel=,
ylabel={Capacity [Mbps]},
scaled y ticks = false,
xmin=0.088,
xmax=3.598,
ymin=0,
ytick pos=right,
yticklabel pos=right,
xticklabels={,,,,,,,,,,},
trim axis right,
width=\hsize,
height=3.8cm,
]
\addplot+[no marks,color=red] table[x=mbits,y=cap] {covert_channel_pc.csv};
\end{axis}
\end{tikzpicture}
    \caption{Desktop setup (Haswell)}
    \label{fig:bitrate_desktop}
  \end{subfigure}

\vspace{0.4cm}

  \begin{subfigure}[b]{\columnwidth}
    \centering
\begin{tikzpicture}[scale=0.9]
\pgfplotsset{every axis legend/.append style={at={(0.5,1.3)},anchor=north,draw=none}}
\begin{axis}[
legend columns=4,
xlabel={Raw bitrate [Mbps]},
ylabel=Bit error probability,
ytick pos=left,
xmin=0.087,
xmax=3.2255,
ymin=0,
ymax=0.5,
scaled y ticks = false,
width=\hsize,
height=3.8cm,
]
\addplot+[no marks, color=blue, dashed] table[x=mbits,y=ber] {covert_channel_server.csv};
\addlegendentry{~Bit error probability \,}
\addplot+[no marks, color=red, draw=none] coordinates {(-5,-5)};
\addlegendentry{~Capacity}
\end{axis}
\begin{axis}[
legend columns=4,
xlabel=,
ylabel={Capacity [Mbps]},
scaled y ticks = false,
xmin=0.087,
xmax=3.2255,
ymin=0,
ytick pos=right,
yticklabel pos=right,
xticklabels={,,,,,,,,,,},
trim axis right,
width=\hsize,
height=3.8cm,
]
\addplot+[no marks,color=red] table[x=mbits,y=cap] {covert_channel_server.csv};
\end{axis}
\end{tikzpicture}
    \caption{Server setup, cross-CPU (Haswell-EP)}
    \label{fig:bitrate_server}
  \end{subfigure}
  
  \caption{Performance of our covert channel implementation (native).}
  \label{fig:bitrate}
\end{figure}

Figure~\ref{fig:bitrate} shows the error rate varying depending on the raw bitrate for the case that both sender and receiver run natively. On our desktop setup (Figure~\ref{fig:bitrate_desktop}), the error probability stays below $1\%$ for bitrates of up to 2\,Mbps. The channel capacity reaches up to 2.1\,Mbps (raw bitrate of 2.4\,Mbps, error probability of $1.8\%$). Beyond this peak, the increasing error probability causes a decrease in the effective capacity.
On our server setup (Figure~\ref{fig:bitrate_server}) the cross-CPU communication achieves 1.2\,Mbps with a $1\%$ error rate. The maximum capacity is 1.6\,Mbps (raw 2.6\,Mbps, 8.7\% error probability).

For the cross-core cross-VM scenario, we deployed two VMs which were configured to use 1\,GB pages for second-stage address translation. We reach a maximum capacity of 309\,kbps (raw 411\,kbps, 4.1\% error probability) on our desktop system. The server setup (cross-CPU cross-VM) performs much better, we achieved a bitrate of 596\,kbps with an error probability of just 0.4\%.

\subsection{Comparison with state of the art}
We compare the bitrate of our DRAM covert channel with the normalized implementation of three cache covert channels by Gruss~\etal\cite{Gruss2016Flush}. For an error rate that is less than 1\%, the covert channel using Prime+Probe obtains 536\,Kbps, the one using Flush+Reload 2.3\,Mbps and the one using Flush+Flush 3.8\,Mbps.
With a capacity of up to 2\,Mbps, our covert channel is within the same order of magnitude of current cache-based channels. However, unlike Flush+Reload and Flush+Flush, it does not require shared memory. Moreover, in contrast to our attack, these cache covert channels do not allow cross-CPU communication.

The work of Irazoqui~\etal\cite{Irazoqui2015cpca} focuses on cross-CPU cache-based side-channel attacks. They did not implement a covert channel, thus we cannot compare our performance with their cache attack. However, their approach also requires shared memory and thus it would not work in our attack setting.

The covert channel by Xiao~\etal\cite{Xiao2013} using memory deduplication achieves up to 90\,bps. However, due to security concerns, memory deduplication has been disabled in many cloud environments. The covert channel of Wu~\etal\cite{Wu2014} using the memory bus achieves 746\,bps with error correction. Our covert channel is therefore three to four orders of magnitude faster than state-of-the-art memory-based covert channels.

\section{A low-noise cross-CPU side channel}\label{sec:sidechannel}
In this section, we present a second DRAMA attack, a highly accurate side-channel attack using DRAM addressing information. We again exploit the row buffer and its behavior similar to a directly-mapped cache. In this attack, the spy and the victim can run on separate CPUs and do not share memory, \ie no access to shared libraries and no page deduplication between VMs.
We mainly consider a local attack scenario where Flush+Reload cache attacks are not applicable due to the lack of shared memory. However, our side-channel attacks can also be applied in a cloud scenario where multiple users on a server and one malicious user spies on other users through this side channel.
The side channel achieves a timing accuracy that is comparable to Flush+Reload and a higher spatial accuracy than Prime+Probe. Thus, it can be used as a highly accurate alternative to Prime+Probe cache attacks in cross-core scenarios without shared memory.

\subsection{Basic concept}

\begin{figure}[tb]
\centering
\includegraphics[width=1.0\linewidth]{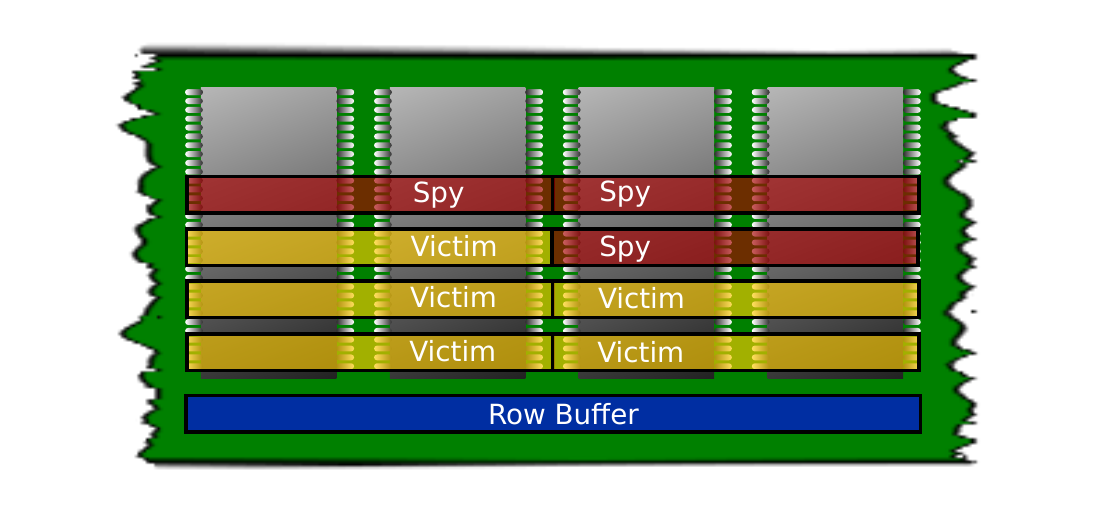}
\caption{Victim and spy have memory allocated in the same DRAM row. By accessing this memory, the spy can determine whether the victim just accessed it.}
\label{fig:row_hit_attack}
\end{figure}

In case of the covert channel, an active sender caused row conflicts. In the side-channel attack, we infer the activity of a victim process by detecting \textit{row hits} and \textit{row conflicts} following our definitions from Section~\ref{sec:definitions}.  For the attack to succeed, spy and victim need to have access to the same row in a bank, as illustrated in Figure~\ref{fig:row_hit_attack}. This is possible without shared memory due to the DRAM addressing functions.

\begin{figure}[tb]
\centering
  \begin{tikzpicture}[scale=0.9]
  \draw (0.2,0.3) node {Page $A$};
  \begin{scope}[transform shape,rotate=-90]
    \draw (0,0) rectangle +(8,0.5);
    \draw[fill=red!20] (0,0) rectangle +(0.2,0.5) node[pos=.5] {\footnotesize 0};
    \draw[fill=blue!20] (0.2,0) rectangle +(0.2,0.5) node[pos=.5] {\footnotesize 1};
    \draw[fill=green!20] (0.4,0) rectangle +(0.2,0.5) node[pos=.5] {\footnotesize 2};
    \draw[fill=yellow!20] (0.6,0) rectangle +(0.2,0.5) node[pos=.5] {\footnotesize 3};
    \draw[fill=black!30] (0.8,0) rectangle +(0.2,0.5) node[pos=.5] {\footnotesize 4};
    \draw[fill=black!10] (1.0,0) rectangle +(0.2,0.5) node[pos=.5] {\footnotesize 5};
    \draw[fill=black!10] (1.2,0) rectangle +(0.2,0.5) node[pos=.5] {\footnotesize 6};
    \draw[fill=black!10] (1.4,0) rectangle +(0.2,0.5) node[pos=.5] {\footnotesize 7};
    \draw[fill=red!20] (1.6,0) rectangle +(0.2,0.5) node[pos=.5] {\footnotesize 0};
    \draw[fill=blue!20] (1.8,0) rectangle +(0.2,0.5) node[pos=.5] {\footnotesize 1};
    \draw[fill=green!20] (2.0,0) rectangle +(0.2,0.5) node[pos=.5] {\footnotesize 2};
    \draw[fill=yellow!20] (2.2,0) rectangle +(0.2,0.5) node[pos=.5] {\footnotesize 3};
    \draw[fill=black!30] (2.4,0) rectangle +(0.2,0.5) node[pos=.5] {\footnotesize 4};
    \draw[fill=black!10] (2.6,0) rectangle +(0.2,0.5) node[pos=.5] {\footnotesize 5};
    \draw[fill=black!10] (2.8,0) rectangle +(0.2,0.5) node[pos=.5] {\footnotesize 6};
    \draw[fill=black!10] (3.0,0) rectangle +(0.2,0.5) node[pos=.5] {\footnotesize 7};
    \draw[fill=red!20] (3.2,0) rectangle +(0.2,0.5) node[pos=.5] {\footnotesize 0};
    \draw[fill=blue!20] (3.4,0) rectangle +(0.2,0.5) node[pos=.5] {\footnotesize 1};
    \draw[fill=green!20] (3.6,0) rectangle +(0.2,0.5) node[pos=.5] {\footnotesize 2};
    \draw[fill=yellow!20] (3.8,0) rectangle +(0.2,0.5) node[pos=.5] {\footnotesize 3};
    \draw[fill=black!30] (4.0,0) rectangle +(0.2,0.5) node[pos=.5] {\footnotesize 4};
    \draw[fill=black!10] (4.2,0) rectangle +(0.2,0.5) node[pos=.5] {\footnotesize 5};
    \draw[fill=black!10] (4.4,0) rectangle +(0.2,0.5) node[pos=.5] {\footnotesize 6};
    \draw[fill=black!10] (4.6,0) rectangle +(0.2,0.5) node[pos=.5] {\footnotesize 7};
    \draw[fill=red!20] (4.8,0) rectangle +(0.2,0.5) node[pos=.5] {\footnotesize 0};
    \draw[fill=blue!20] (5.0,0) rectangle +(0.2,0.5) node[pos=.5] {\footnotesize 1};
    \draw[fill=green!20] (5.2,0) rectangle +(0.2,0.5) node[pos=.5] {\footnotesize 2};
    \draw[fill=yellow!20] (5.4,0) rectangle +(0.2,0.5) node[pos=.5] {\footnotesize 3};
    \draw[fill=black!30] (5.6,0) rectangle +(0.2,0.5) node[pos=.5] {\footnotesize 4};
    \draw[fill=black!10] (5.8,0) rectangle +(0.2,0.5) node[pos=.5] {\footnotesize 5};
    \draw[fill=black!10] (6.0,0) rectangle +(0.2,0.5) node[pos=.5] {\footnotesize 6};
    \draw[fill=black!10] (6.2,0) rectangle +(0.2,0.5) node[pos=.5] {\footnotesize 7};
    \draw[fill=red!20] (6.4,0) rectangle +(0.2,0.5) node[pos=.5] {\footnotesize 0};
    \draw[fill=blue!20] (6.6,0) rectangle +(0.2,0.5) node[pos=.5] {\footnotesize 1};
    \draw[fill=green!20] (6.8,0) rectangle +(0.2,0.5) node[pos=.5] {\footnotesize 2};
    \draw[fill=yellow!20] (7.0,0) rectangle +(0.2,0.5) node[pos=.5] {\footnotesize 3};
    \draw[fill=black!30] (7.2,0) rectangle +(0.2,0.5) node[pos=.5] {\footnotesize 4};
    \draw[fill=black!10] (7.4,0) rectangle +(0.2,0.5) node[pos=.5] {\footnotesize 5};
    \draw[fill=black!10] (7.6,0) rectangle +(0.2,0.5) node[pos=.5] {\footnotesize 6};
    \draw[fill=black!10] (7.8,0) rectangle +(0.2,0.5) node[pos=.5] {\footnotesize 7};
\draw [decorate,decoration={brace,amplitude=10pt}]
(8,-0.1) -- (0,-0.1) node [midway,yshift=-0.6cm] 
{$64 \cdot 64$ bytes (4\,KB page)};
    \draw[fill=white,white,in=270,out=90] (3.85,0) to node{} +(0.05,0.125) to node[rotate=90,yshift=-0.13cm]{
    } +(-0.1,0.25) to node{} +(+0.05,0.125) to node{} +(+0.28,0) to node{} +(-0.05,-0.125) to node{} +(+0.1,-0.25) to node{} +(-0.05,-0.125) to node{} (4.15,0);
    \draw[very thick,white] (3.86,0) -- (4.14,0);
    \draw[very thick,white] (3.86,0.5) -- (4.14,0.5);
    \draw[in=270,out=90] (3.85,0) to node{} +(0.05,0.125) to node[rotate=90,yshift=-0.13cm]{
    } +(-0.1,0.25) to node{} +(+0.05,0.125);
    \draw[in=270,out=90] (4.15,0) to node{} +(0.05,0.125) to node{} +(-0.1,0.25) to node{} +(+0.05,0.125);

  \end{scope}

  \draw[red!60] (0.5,-0.1) -- +(0.2,0) -- (0.7,-1.2);
  \draw[red!60] (0.5,-1.7) -- +(0.2,0) -- (0.7,-1.2);
  \draw[red!60] (0.5,-3.3) -- +(0.2,0) -- (0.7,-1.2);
  \draw[red!60] (0.5,-4.9) -- +(0.2,0) -- (0.7,-1.2);
  \draw[red!60] (0.5,-6.5) -- +(0.2,0) -- (0.7,-1.2);
  \draw[red!60] (0.7,-1.2) -- (2.0,-1.2) -- (2.0,-0.8);
  \draw[red!60] (0.7,-1.2) -- (2.1,-1.2) -- (2.1,-0.8);
  \draw[red!60] (0.7,-1.2) -- (2.2,-1.2) -- (2.2,-0.8);
  \draw[red!60] (0.7,-1.2) -- (2.3,-1.2) -- (2.3,-0.8);
  \draw[red!60] (0.7,-1.2) -- (2.4,-1.2) -- (2.4,-0.8);

  \draw[blue!60] (0.5,-0.3) -- +(0.4,0) -- (0.9,-2.7);
  \draw[blue!60] (0.5,-1.9) -- +(0.4,0) -- (0.9,-2.7);
  \draw[blue!60] (0.5,-3.5) -- +(0.4,0) -- (0.9,-2.7);
  \draw[blue!60] (0.5,-5.1) -- +(0.4,0) -- (0.9,-2.7);
  \draw[blue!60] (0.5,-6.7) -- +(0.4,0) -- (0.9,-2.7);
  \draw[blue!60] (0.9,-2.7) -- (2.0,-2.7) -- (2.0,-2.3);
  \draw[blue!60] (0.9,-2.7) -- (2.1,-2.7) -- (2.1,-2.3);
  \draw[blue!60] (0.9,-2.7) -- (2.2,-2.7) -- (2.2,-2.3);
  \draw[blue!60] (0.9,-2.7) -- (2.3,-2.7) -- (2.3,-2.3);
  \draw[blue!60] (0.9,-2.7) -- (2.4,-2.7) -- (2.4,-2.3);
    
  \draw[black!40!green!60] (0.5,-0.5) -- +(0.6,0) -- (1.1,-4.2);
  \draw[black!40!green!60] (0.5,-2.1) -- +(0.6,0) -- (1.1,-4.2);
  \draw[black!40!green!60] (0.5,-3.7) -- +(0.6,0) -- (1.1,-4.2);
  \draw[black!40!green!60] (0.5,-5.3) -- +(0.6,0) -- (1.1,-4.2);
  \draw[black!40!green!60] (0.5,-6.9) -- +(0.6,0) -- (1.1,-4.2);
  \draw[black!40!green!60] (1.1,-4.2) -- (2.0,-4.2) -- (2.0,-3.8);
  \draw[black!40!green!60] (1.1,-4.2) -- (2.1,-4.2) -- (2.1,-3.8);
  \draw[black!40!green!60] (1.1,-4.2) -- (2.2,-4.2) -- (2.2,-3.8);
  \draw[black!40!green!60] (1.1,-4.2) -- (2.3,-4.2) -- (2.3,-3.8);
  \draw[black!40!green!60] (1.1,-4.2) -- (2.4,-4.2) -- (2.4,-3.8);
      
  \draw[black!40!yellow!60] (0.5,-0.7) -- +(0.8,0) -- (1.3,-5.6);
  \draw[black!40!yellow!60] (0.5,-2.3) -- +(0.8,0) -- (1.3,-5.6);
  \draw[black!40!yellow!60] (0.5,-5.5) -- +(0.8,0) -- (1.3,-5.6);
  \draw[black!40!yellow!60] (0.5,-7.1) -- +(0.8,0) -- (1.3,-5.6);
  \draw[black!40!yellow!60] (1.3,-5.6) -- (2.0,-5.6) -- (2.0,-5.3);
  \draw[black!40!yellow!60] (1.3,-5.6) -- (2.1,-5.6) -- (2.1,-5.3);
  \draw[black!40!yellow!60] (1.3,-5.6) -- (2.2,-5.6) -- (2.2,-5.3);
  \draw[black!40!yellow!60] (1.3,-5.6) -- (2.3,-5.6) -- (2.3,-5.3);
  \draw[black!40!yellow!60] (1.3,-5.6) -- (2.4,-5.6) -- (2.4,-5.3);

  \draw[black!60] (0.5,-0.9) -- +(1.0,0) -- (1.5,-7.2);
  \draw[black!60] (0.5,-2.5) -- +(1.0,0) -- (1.5,-7.2);
  \draw[black!60] (0.5,-5.7) -- +(1.0,0) -- (1.5,-7.2);
  \draw[black!60] (0.5,-7.3) -- +(1.0,0) -- (1.5,-7.2);
  \draw[black!60] (1.5,-7.2) -- (2.0,-7.2) -- (2.0,-6.8);
  \draw[black!60] (1.5,-7.2) -- (2.1,-7.2) -- (2.1,-6.8);
  \draw[black!60] (1.5,-7.2) -- (2.2,-7.2) -- (2.2,-6.8);
  \draw[black!60] (1.5,-7.2) -- (2.3,-7.2) -- (2.3,-6.8);
  \draw[black!60] (1.5,-7.2) -- (2.4,-7.2) -- (2.4,-6.8);
    
    \draw (4.75,-4) node[opacity=0.3] {\includegraphics[trim=50 40 50 22,clip,width=0.68\hsize,height=7cm]{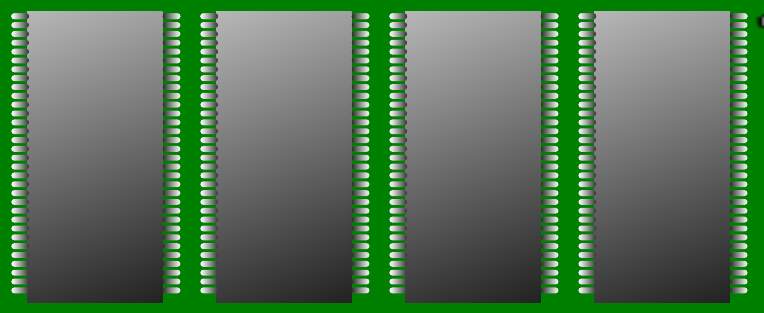}};
  \begin{scope}[shift={(1.93,-0.87)},transform shape,scale=0.7]
    \draw (0,0) rectangle +(8,0.5);
    \draw[fill=red!20] (0,0) rectangle +(0.2,0.5) node[pos=.5] {\footnotesize $A$};
    \draw[fill=red!20] (0.2,0) rectangle +(0.2,0.5) node[pos=.5] {\footnotesize $A$};
    \draw[fill=red!20] (0.4,0) rectangle +(0.2,0.5) node[pos=.5] {\footnotesize $A$};
    \draw[fill=red!20] (0.6,0) rectangle +(0.2,0.5) node[pos=.5] {\footnotesize $A$};
    \draw[fill=red!20] (0.8,0) rectangle +(0.2,0.5) node[pos=.5] {\footnotesize $A$};
    \draw[fill=red!20] (1.0,0) rectangle +(0.2,0.5) node[pos=.5] {\footnotesize $A$};
    \draw[fill=red!20] (1.2,0) rectangle +(0.2,0.5) node[pos=.5] {\footnotesize $A$};
    \draw[fill=red!20] (1.4,0) rectangle +(0.2,0.5) node[pos=.5] {\footnotesize $A$};
    \draw[fill=black!10] (1.6,0) rectangle +(0.2,0.5) node[pos=.5] {\footnotesize $B$};
    \draw[fill=black!10] (1.8,0) rectangle +(0.2,0.5) node[pos=.5] {\footnotesize $B$};
    \draw[fill=black!10] (2.0,0) rectangle +(0.2,0.5) node[pos=.5] {\footnotesize $B$};
    \draw[fill=black!10] (2.2,0) rectangle +(0.2,0.5) node[pos=.5] {\footnotesize $B$};
    \draw[fill=black!10] (2.4,0) rectangle +(0.2,0.5) node[pos=.5] {\footnotesize $B$};
    \draw[fill=black!10] (2.6,0) rectangle +(0.2,0.5) node[pos=.5] {\footnotesize $B$};
    \draw[fill=black!10] (2.8,0) rectangle +(0.2,0.5) node[pos=.5] {\footnotesize $B$};
    \draw[fill=black!10] (3.0,0) rectangle +(0.2,0.5) node[pos=.5] {\footnotesize $B$};
    \draw[fill=black!10] (3.2,0) rectangle +(0.2,0.5) node[pos=.5] {\footnotesize $C$};
    \draw[fill=black!10] (3.4,0) rectangle +(0.2,0.5) node[pos=.5] {\footnotesize $C$};
    \draw[fill=black!10] (3.6,0) rectangle +(0.2,0.5) node[pos=.5] {\footnotesize $C$};
    \draw[fill=black!10] (3.8,0) rectangle +(0.2,0.5) node[pos=.5] {\footnotesize $C$};
    \draw[fill=black!10] (4.0,0) rectangle +(0.2,0.5) node[pos=.5] {\footnotesize $N$};
    \draw[fill=black!10] (4.2,0) rectangle +(0.2,0.5) node[pos=.5] {\footnotesize $N$};
    \draw[fill=black!10] (4.4,0) rectangle +(0.2,0.5) node[pos=.5] {\footnotesize $N$};
    \draw[fill=black!10] (4.6,0) rectangle +(0.2,0.5) node[pos=.5] {\footnotesize $N$};
    \draw[fill=black!10] (4.8,0) rectangle +(0.2,0.5) node[pos=.5] {\footnotesize $O$};
    \draw[fill=black!10] (5.0,0) rectangle +(0.2,0.5) node[pos=.5] {\footnotesize $O$};
    \draw[fill=black!10] (5.2,0) rectangle +(0.2,0.5) node[pos=.5] {\footnotesize $O$};
    \draw[fill=black!10] (5.4,0) rectangle +(0.2,0.5) node[pos=.5] {\footnotesize $O$};
    \draw[fill=black!10] (5.6,0) rectangle +(0.2,0.5) node[pos=.5] {\footnotesize $O$};
    \draw[fill=black!10] (5.8,0) rectangle +(0.2,0.5) node[pos=.5] {\footnotesize $O$};
    \draw[fill=black!10] (6.0,0) rectangle +(0.2,0.5) node[pos=.5] {\footnotesize $O$};
    \draw[fill=black!10] (6.2,0) rectangle +(0.2,0.5) node[pos=.5] {\footnotesize $O$};
    \draw[fill=black!10] (6.4,0) rectangle +(0.2,0.5) node[pos=.5] {\footnotesize $P$};
    \draw[fill=black!10] (6.6,0) rectangle +(0.2,0.5) node[pos=.5] {\footnotesize $P$};
    \draw[fill=black!10] (6.8,0) rectangle +(0.2,0.5) node[pos=.5] {\footnotesize $P$};
    \draw[fill=black!10] (7.0,0) rectangle +(0.2,0.5) node[pos=.5] {\footnotesize $P$};
    \draw[fill=black!10] (7.2,0) rectangle +(0.2,0.5) node[pos=.5] {\footnotesize $P$};
    \draw[fill=black!10] (7.4,0) rectangle +(0.2,0.5) node[pos=.5] {\footnotesize $P$};
    \draw[fill=black!10] (7.6,0) rectangle +(0.2,0.5) node[pos=.5] {\footnotesize $P$};
    \draw[fill=black!10] (7.8,0) rectangle +(0.2,0.5) node[pos=.5] {\footnotesize $P$};
\draw [decorate,decoration={brace,amplitude=10pt}]
(0,0.6) -- (8,0.6) node [transform shape,scale=1.42,midway,yshift=0.7cm] 
{$128 \cdot 64$ bytes (8\,KB DRAM row)};
\draw (4,-0.25) node[transform shape,scale=1.42] {Row in bank 0};
    \draw[fill=green!50!black!30,green!50!black!30,in=270,out=90] (3.85,0) to node{} +(0.05,0.125) to node[rotate=90,yshift=-0.13cm]{
    } +(-0.1,0.25) to node{} +(+0.05,0.125) to node{} +(+0.28,0) to node{} +(-0.05,-0.125) to node{} +(+0.1,-0.25) to node{} +(-0.05,-0.125) to node{} (4.15,0);
    \draw[very thick,green!50!black!30] (3.86,0) -- (4.14,0);
    \draw[very thick,green!50!black!30] (3.86,0.5) -- (4.14,0.5);
    \draw[in=270,out=90] (3.85,0) to node{} +(0.05,0.125) to node[rotate=90,yshift=-0.13cm]{
    } +(-0.1,0.25) to node{} +(+0.05,0.125);
    \draw[in=270,out=90] (4.15,0) to node{} +(0.05,0.125) to node{} +(-0.1,0.25) to node{} +(+0.05,0.125);

  \end{scope}
  \begin{scope}[shift={(1.93,-2.37)},transform shape,scale=0.7]
    \draw (0,0) rectangle +(8,0.5);
    \draw[fill=blue!20] (0,0) rectangle +(0.2,0.5) node[pos=.5] {\footnotesize $A$};
    \draw[fill=blue!20] (0.2,0) rectangle +(0.2,0.5) node[pos=.5] {\footnotesize $A$};
    \draw[fill=blue!20] (0.4,0) rectangle +(0.2,0.5) node[pos=.5] {\footnotesize $A$};
    \draw[fill=blue!20] (0.6,0) rectangle +(0.2,0.5) node[pos=.5] {\footnotesize $A$};
    \draw[fill=blue!20] (0.8,0) rectangle +(0.2,0.5) node[pos=.5] {\footnotesize $A$};
    \draw[fill=blue!20] (1.0,0) rectangle +(0.2,0.5) node[pos=.5] {\footnotesize $A$};
    \draw[fill=blue!20] (1.2,0) rectangle +(0.2,0.5) node[pos=.5] {\footnotesize $A$};
    \draw[fill=blue!20] (1.4,0) rectangle +(0.2,0.5) node[pos=.5] {\footnotesize $A$};
    \draw[fill=black!10] (1.6,0) rectangle +(0.2,0.5) node[pos=.5] {\footnotesize $B$};
    \draw[fill=black!10] (1.8,0) rectangle +(0.2,0.5) node[pos=.5] {\footnotesize $B$};
    \draw[fill=black!10] (2.0,0) rectangle +(0.2,0.5) node[pos=.5] {\footnotesize $B$};
    \draw[fill=black!10] (2.2,0) rectangle +(0.2,0.5) node[pos=.5] {\footnotesize $B$};
    \draw[fill=black!10] (2.4,0) rectangle +(0.2,0.5) node[pos=.5] {\footnotesize $B$};
    \draw[fill=black!10] (2.6,0) rectangle +(0.2,0.5) node[pos=.5] {\footnotesize $B$};
    \draw[fill=black!10] (2.8,0) rectangle +(0.2,0.5) node[pos=.5] {\footnotesize $B$};
    \draw[fill=black!10] (3.0,0) rectangle +(0.2,0.5) node[pos=.5] {\footnotesize $B$};
    \draw[fill=black!10] (3.2,0) rectangle +(0.2,0.5) node[pos=.5] {\footnotesize $C$};
    \draw[fill=black!10] (3.4,0) rectangle +(0.2,0.5) node[pos=.5] {\footnotesize $C$};
    \draw[fill=black!10] (3.6,0) rectangle +(0.2,0.5) node[pos=.5] {\footnotesize $C$};
    \draw[fill=black!10] (3.8,0) rectangle +(0.2,0.5) node[pos=.5] {\footnotesize $C$};
    \draw[fill=black!10] (4.0,0) rectangle +(0.2,0.5) node[pos=.5] {\footnotesize $N$};
    \draw[fill=black!10] (4.2,0) rectangle +(0.2,0.5) node[pos=.5] {\footnotesize $N$};
    \draw[fill=black!10] (4.4,0) rectangle +(0.2,0.5) node[pos=.5] {\footnotesize $N$};
    \draw[fill=black!10] (4.6,0) rectangle +(0.2,0.5) node[pos=.5] {\footnotesize $N$};
    \draw[fill=black!10] (4.8,0) rectangle +(0.2,0.5) node[pos=.5] {\footnotesize $O$};
    \draw[fill=black!10] (5.0,0) rectangle +(0.2,0.5) node[pos=.5] {\footnotesize $O$};
    \draw[fill=black!10] (5.2,0) rectangle +(0.2,0.5) node[pos=.5] {\footnotesize $O$};
    \draw[fill=black!10] (5.4,0) rectangle +(0.2,0.5) node[pos=.5] {\footnotesize $O$};
    \draw[fill=black!10] (5.6,0) rectangle +(0.2,0.5) node[pos=.5] {\footnotesize $O$};
    \draw[fill=black!10] (5.8,0) rectangle +(0.2,0.5) node[pos=.5] {\footnotesize $O$};
    \draw[fill=black!10] (6.0,0) rectangle +(0.2,0.5) node[pos=.5] {\footnotesize $O$};
    \draw[fill=black!10] (6.2,0) rectangle +(0.2,0.5) node[pos=.5] {\footnotesize $O$};
    \draw[fill=black!10] (6.4,0) rectangle +(0.2,0.5) node[pos=.5] {\footnotesize $P$};
    \draw[fill=black!10] (6.6,0) rectangle +(0.2,0.5) node[pos=.5] {\footnotesize $P$};
    \draw[fill=black!10] (6.8,0) rectangle +(0.2,0.5) node[pos=.5] {\footnotesize $P$};
    \draw[fill=black!10] (7.0,0) rectangle +(0.2,0.5) node[pos=.5] {\footnotesize $P$};
    \draw[fill=black!10] (7.2,0) rectangle +(0.2,0.5) node[pos=.5] {\footnotesize $P$};
    \draw[fill=black!10] (7.4,0) rectangle +(0.2,0.5) node[pos=.5] {\footnotesize $P$};
    \draw[fill=black!10] (7.6,0) rectangle +(0.2,0.5) node[pos=.5] {\footnotesize $P$};
    \draw[fill=black!10] (7.8,0) rectangle +(0.2,0.5) node[pos=.5] {\footnotesize $P$};
\draw (4,-0.25) node[transform shape,scale=1.42] {Row in bank 1};
    \draw[fill=green!50!black!30,green!50!black!30,in=270,out=90] (3.85,0) to node{} +(0.05,0.125) to node[rotate=90,yshift=-0.13cm]{
    } +(-0.1,0.25) to node{} +(+0.05,0.125) to node{} +(+0.28,0) to node{} +(-0.05,-0.125) to node{} +(+0.1,-0.25) to node{} +(-0.05,-0.125) to node{} (4.15,0);
    \draw[very thick,green!50!black!30] (3.86,0) -- (4.14,0);
    \draw[very thick,green!50!black!30] (3.86,0.5) -- (4.14,0.5);
    \draw[in=270,out=90] (3.85,0) to node{} +(0.05,0.125) to node[rotate=90,yshift=-0.13cm]{
    } +(-0.1,0.25) to node{} +(+0.05,0.125);
    \draw[in=270,out=90] (4.15,0) to node{} +(0.05,0.125) to node{} +(-0.1,0.25) to node{} +(+0.05,0.125);

  \end{scope}
  
    \begin{scope}[shift={(1.93,-3.87)},transform shape,scale=0.7]
    \draw (0,0) rectangle +(8,0.5);
    \draw[fill=green!20] (0,0) rectangle +(0.2,0.5) node[pos=.5] {\footnotesize $A$};
    \draw[fill=green!20] (0.2,0) rectangle +(0.2,0.5) node[pos=.5] {\footnotesize $A$};
    \draw[fill=green!20] (0.4,0) rectangle +(0.2,0.5) node[pos=.5] {\footnotesize $A$};
    \draw[fill=green!20] (0.6,0) rectangle +(0.2,0.5) node[pos=.5] {\footnotesize $A$};
    \draw[fill=green!20] (0.8,0) rectangle +(0.2,0.5) node[pos=.5] {\footnotesize $A$};
    \draw[fill=green!20] (1.0,0) rectangle +(0.2,0.5) node[pos=.5] {\footnotesize $A$};
    \draw[fill=green!20] (1.2,0) rectangle +(0.2,0.5) node[pos=.5] {\footnotesize $A$};
    \draw[fill=green!20] (1.4,0) rectangle +(0.2,0.5) node[pos=.5] {\footnotesize $A$};
    \draw[fill=black!10] (1.6,0) rectangle +(0.2,0.5) node[pos=.5] {\footnotesize $B$};
    \draw[fill=black!10] (1.8,0) rectangle +(0.2,0.5) node[pos=.5] {\footnotesize $B$};
    \draw[fill=black!10] (2.0,0) rectangle +(0.2,0.5) node[pos=.5] {\footnotesize $B$};
    \draw[fill=black!10] (2.2,0) rectangle +(0.2,0.5) node[pos=.5] {\footnotesize $B$};
    \draw[fill=black!10] (2.4,0) rectangle +(0.2,0.5) node[pos=.5] {\footnotesize $B$};
    \draw[fill=black!10] (2.6,0) rectangle +(0.2,0.5) node[pos=.5] {\footnotesize $B$};
    \draw[fill=black!10] (2.8,0) rectangle +(0.2,0.5) node[pos=.5] {\footnotesize $B$};
    \draw[fill=black!10] (3.0,0) rectangle +(0.2,0.5) node[pos=.5] {\footnotesize $B$};
    \draw[fill=black!10] (3.2,0) rectangle +(0.2,0.5) node[pos=.5] {\footnotesize $C$};
    \draw[fill=black!10] (3.4,0) rectangle +(0.2,0.5) node[pos=.5] {\footnotesize $C$};
    \draw[fill=black!10] (3.6,0) rectangle +(0.2,0.5) node[pos=.5] {\footnotesize $C$};
    \draw[fill=black!10] (3.8,0) rectangle +(0.2,0.5) node[pos=.5] {\footnotesize $C$};
    \draw[fill=black!10] (4.0,0) rectangle +(0.2,0.5) node[pos=.5] {\footnotesize $N$};
    \draw[fill=black!10] (4.2,0) rectangle +(0.2,0.5) node[pos=.5] {\footnotesize $N$};
    \draw[fill=black!10] (4.4,0) rectangle +(0.2,0.5) node[pos=.5] {\footnotesize $N$};
    \draw[fill=black!10] (4.6,0) rectangle +(0.2,0.5) node[pos=.5] {\footnotesize $N$};
    \draw[fill=black!10] (4.8,0) rectangle +(0.2,0.5) node[pos=.5] {\footnotesize $O$};
    \draw[fill=black!10] (5.0,0) rectangle +(0.2,0.5) node[pos=.5] {\footnotesize $O$};
    \draw[fill=black!10] (5.2,0) rectangle +(0.2,0.5) node[pos=.5] {\footnotesize $O$};
    \draw[fill=black!10] (5.4,0) rectangle +(0.2,0.5) node[pos=.5] {\footnotesize $O$};
    \draw[fill=black!10] (5.6,0) rectangle +(0.2,0.5) node[pos=.5] {\footnotesize $O$};
    \draw[fill=black!10] (5.8,0) rectangle +(0.2,0.5) node[pos=.5] {\footnotesize $O$};
    \draw[fill=black!10] (6.0,0) rectangle +(0.2,0.5) node[pos=.5] {\footnotesize $O$};
    \draw[fill=black!10] (6.2,0) rectangle +(0.2,0.5) node[pos=.5] {\footnotesize $O$};
    \draw[fill=black!10] (6.4,0) rectangle +(0.2,0.5) node[pos=.5] {\footnotesize $P$};
    \draw[fill=black!10] (6.6,0) rectangle +(0.2,0.5) node[pos=.5] {\footnotesize $P$};
    \draw[fill=black!10] (6.8,0) rectangle +(0.2,0.5) node[pos=.5] {\footnotesize $P$};
    \draw[fill=black!10] (7.0,0) rectangle +(0.2,0.5) node[pos=.5] {\footnotesize $P$};
    \draw[fill=black!10] (7.2,0) rectangle +(0.2,0.5) node[pos=.5] {\footnotesize $P$};
    \draw[fill=black!10] (7.4,0) rectangle +(0.2,0.5) node[pos=.5] {\footnotesize $P$};
    \draw[fill=black!10] (7.6,0) rectangle +(0.2,0.5) node[pos=.5] {\footnotesize $P$};
    \draw[fill=black!10] (7.8,0) rectangle +(0.2,0.5) node[pos=.5] {\footnotesize $P$};
\draw (4,-0.25) node[transform shape,scale=1.42] {Row in bank 2};
    \draw[fill=green!50!black!30,green!50!black!30,in=270,out=90] (3.85,0) to node{} +(0.05,0.125) to node[rotate=90,yshift=-0.13cm]{
    } +(-0.1,0.25) to node{} +(+0.05,0.125) to node{} +(+0.28,0) to node{} +(-0.05,-0.125) to node{} +(+0.1,-0.25) to node{} +(-0.05,-0.125) to node{} (4.15,0);
    \draw[very thick,green!50!black!30] (3.86,0) -- (4.14,0);
    \draw[very thick,green!50!black!30] (3.86,0.5) -- (4.14,0.5);
    \draw[in=270,out=90] (3.85,0) to node{} +(0.05,0.125) to node[rotate=90,yshift=-0.13cm]{
    } +(-0.1,0.25) to node{} +(+0.05,0.125);
    \draw[in=270,out=90] (4.15,0) to node{} +(0.05,0.125) to node{} +(-0.1,0.25) to node{} +(+0.05,0.125);

  \end{scope}
  
    \begin{scope}[shift={(1.93,-5.37)},transform shape,scale=0.7]
    \draw (0,0) rectangle +(8,0.5);
    \draw[fill=yellow!20] (0,0) rectangle +(0.2,0.5) node[pos=.5] {\footnotesize $A$};
    \draw[fill=yellow!20] (0.2,0) rectangle +(0.2,0.5) node[pos=.5] {\footnotesize $A$};
    \draw[fill=yellow!20] (0.4,0) rectangle +(0.2,0.5) node[pos=.5] {\footnotesize $A$};
    \draw[fill=yellow!20] (0.6,0) rectangle +(0.2,0.5) node[pos=.5] {\footnotesize $A$};
    \draw[fill=yellow!20] (0.8,0) rectangle +(0.2,0.5) node[pos=.5] {\footnotesize $A$};
    \draw[fill=yellow!20] (1.0,0) rectangle +(0.2,0.5) node[pos=.5] {\footnotesize $A$};
    \draw[fill=yellow!20] (1.2,0) rectangle +(0.2,0.5) node[pos=.5] {\footnotesize $A$};
    \draw[fill=yellow!20] (1.4,0) rectangle +(0.2,0.5) node[pos=.5] {\footnotesize $A$};
    \draw[fill=black!10] (1.6,0) rectangle +(0.2,0.5) node[pos=.5] {\footnotesize $B$};
    \draw[fill=black!10] (1.8,0) rectangle +(0.2,0.5) node[pos=.5] {\footnotesize $B$};
    \draw[fill=black!10] (2.0,0) rectangle +(0.2,0.5) node[pos=.5] {\footnotesize $B$};
    \draw[fill=black!10] (2.2,0) rectangle +(0.2,0.5) node[pos=.5] {\footnotesize $B$};
    \draw[fill=black!10] (2.4,0) rectangle +(0.2,0.5) node[pos=.5] {\footnotesize $B$};
    \draw[fill=black!10] (2.6,0) rectangle +(0.2,0.5) node[pos=.5] {\footnotesize $B$};
    \draw[fill=black!10] (2.8,0) rectangle +(0.2,0.5) node[pos=.5] {\footnotesize $B$};
    \draw[fill=black!10] (3.0,0) rectangle +(0.2,0.5) node[pos=.5] {\footnotesize $B$};
    \draw[fill=black!10] (3.2,0) rectangle +(0.2,0.5) node[pos=.5] {\footnotesize $C$};
    \draw[fill=black!10] (3.4,0) rectangle +(0.2,0.5) node[pos=.5] {\footnotesize $C$};
    \draw[fill=black!10] (3.6,0) rectangle +(0.2,0.5) node[pos=.5] {\footnotesize $C$};
    \draw[fill=black!10] (3.8,0) rectangle +(0.2,0.5) node[pos=.5] {\footnotesize $C$};
    \draw[fill=black!10] (4.0,0) rectangle +(0.2,0.5) node[pos=.5] {\footnotesize $N$};
    \draw[fill=black!10] (4.2,0) rectangle +(0.2,0.5) node[pos=.5] {\footnotesize $N$};
    \draw[fill=black!10] (4.4,0) rectangle +(0.2,0.5) node[pos=.5] {\footnotesize $N$};
    \draw[fill=black!10] (4.6,0) rectangle +(0.2,0.5) node[pos=.5] {\footnotesize $N$};
    \draw[fill=black!10] (4.8,0) rectangle +(0.2,0.5) node[pos=.5] {\footnotesize $O$};
    \draw[fill=black!10] (5.0,0) rectangle +(0.2,0.5) node[pos=.5] {\footnotesize $O$};
    \draw[fill=black!10] (5.2,0) rectangle +(0.2,0.5) node[pos=.5] {\footnotesize $O$};
    \draw[fill=black!10] (5.4,0) rectangle +(0.2,0.5) node[pos=.5] {\footnotesize $O$};
    \draw[fill=black!10] (5.6,0) rectangle +(0.2,0.5) node[pos=.5] {\footnotesize $O$};
    \draw[fill=black!10] (5.8,0) rectangle +(0.2,0.5) node[pos=.5] {\footnotesize $O$};
    \draw[fill=black!10] (6.0,0) rectangle +(0.2,0.5) node[pos=.5] {\footnotesize $O$};
    \draw[fill=black!10] (6.2,0) rectangle +(0.2,0.5) node[pos=.5] {\footnotesize $O$};
    \draw[fill=black!10] (6.4,0) rectangle +(0.2,0.5) node[pos=.5] {\footnotesize $P$};
    \draw[fill=black!10] (6.6,0) rectangle +(0.2,0.5) node[pos=.5] {\footnotesize $P$};
    \draw[fill=black!10] (6.8,0) rectangle +(0.2,0.5) node[pos=.5] {\footnotesize $P$};
    \draw[fill=black!10] (7.0,0) rectangle +(0.2,0.5) node[pos=.5] {\footnotesize $P$};
    \draw[fill=black!10] (7.2,0) rectangle +(0.2,0.5) node[pos=.5] {\footnotesize $P$};
    \draw[fill=black!10] (7.4,0) rectangle +(0.2,0.5) node[pos=.5] {\footnotesize $P$};
    \draw[fill=black!10] (7.6,0) rectangle +(0.2,0.5) node[pos=.5] {\footnotesize $P$};
    \draw[fill=black!10] (7.8,0) rectangle +(0.2,0.5) node[pos=.5] {\footnotesize $P$};
\draw (4,-0.25) node[transform shape,scale=1.42] {Row in bank 3};
    \draw[fill=green!50!black!30,green!50!black!30,in=270,out=90] (3.85,0) to node{} +(0.05,0.125) to node[rotate=90,yshift=-0.13cm]{
    } +(-0.1,0.25) to node{} +(+0.05,0.125) to node{} +(+0.28,0) to node{} +(-0.05,-0.125) to node{} +(+0.1,-0.25) to node{} +(-0.05,-0.125) to node{} (4.15,0);
    \draw[very thick,green!50!black!30] (3.86,0) -- (4.14,0);
    \draw[very thick,green!50!black!30] (3.86,0.5) -- (4.14,0.5);
    \draw[in=270,out=90] (3.85,0) to node{} +(0.05,0.125) to node[rotate=90,yshift=-0.13cm]{
    } +(-0.1,0.25) to node{} +(+0.05,0.125);
    \draw[in=270,out=90] (4.15,0) to node{} +(0.05,0.125) to node{} +(-0.1,0.25) to node{} +(+0.05,0.125);

  \end{scope}
  \begin{scope}[shift={(1.93,-6.87)},transform shape,scale=0.7]
    \draw (0,0) rectangle +(8,0.5);
    \draw[fill=black!30] (0,0) rectangle +(0.2,0.5) node[pos=.5] {\footnotesize $A$};
    \draw[fill=black!30] (0.2,0) rectangle +(0.2,0.5) node[pos=.5] {\footnotesize $A$};
    \draw[fill=black!30] (0.4,0) rectangle +(0.2,0.5) node[pos=.5] {\footnotesize $A$};
    \draw[fill=black!30] (0.6,0) rectangle +(0.2,0.5) node[pos=.5] {\footnotesize $A$};
    \draw[fill=black!30] (0.8,0) rectangle +(0.2,0.5) node[pos=.5] {\footnotesize $A$};
    \draw[fill=black!30] (1.0,0) rectangle +(0.2,0.5) node[pos=.5] {\footnotesize $A$};
    \draw[fill=black!30] (1.2,0) rectangle +(0.2,0.5) node[pos=.5] {\footnotesize $A$};
    \draw[fill=black!30] (1.4,0) rectangle +(0.2,0.5) node[pos=.5] {\footnotesize $A$};
    \draw[fill=black!10] (1.6,0) rectangle +(0.2,0.5) node[pos=.5] {\footnotesize $B$};
    \draw[fill=black!10] (1.8,0) rectangle +(0.2,0.5) node[pos=.5] {\footnotesize $B$};
    \draw[fill=black!10] (2.0,0) rectangle +(0.2,0.5) node[pos=.5] {\footnotesize $B$};
    \draw[fill=black!10] (2.2,0) rectangle +(0.2,0.5) node[pos=.5] {\footnotesize $B$};
    \draw[fill=black!10] (2.4,0) rectangle +(0.2,0.5) node[pos=.5] {\footnotesize $B$};
    \draw[fill=black!10] (2.6,0) rectangle +(0.2,0.5) node[pos=.5] {\footnotesize $B$};
    \draw[fill=black!10] (2.8,0) rectangle +(0.2,0.5) node[pos=.5] {\footnotesize $B$};
    \draw[fill=black!10] (3.0,0) rectangle +(0.2,0.5) node[pos=.5] {\footnotesize $B$};
    \draw[fill=black!10] (3.2,0) rectangle +(0.2,0.5) node[pos=.5] {\footnotesize $C$};
    \draw[fill=black!10] (3.4,0) rectangle +(0.2,0.5) node[pos=.5] {\footnotesize $C$};
    \draw[fill=black!10] (3.6,0) rectangle +(0.2,0.5) node[pos=.5] {\footnotesize $C$};
    \draw[fill=black!10] (3.8,0) rectangle +(0.2,0.5) node[pos=.5] {\footnotesize $C$};
    \draw[fill=black!10] (4.0,0) rectangle +(0.2,0.5) node[pos=.5] {\footnotesize $N$};
    \draw[fill=black!10] (4.2,0) rectangle +(0.2,0.5) node[pos=.5] {\footnotesize $N$};
    \draw[fill=black!10] (4.4,0) rectangle +(0.2,0.5) node[pos=.5] {\footnotesize $N$};
    \draw[fill=black!10] (4.6,0) rectangle +(0.2,0.5) node[pos=.5] {\footnotesize $N$};
    \draw[fill=black!10] (4.8,0) rectangle +(0.2,0.5) node[pos=.5] {\footnotesize $O$};
    \draw[fill=black!10] (5.0,0) rectangle +(0.2,0.5) node[pos=.5] {\footnotesize $O$};
    \draw[fill=black!10] (5.2,0) rectangle +(0.2,0.5) node[pos=.5] {\footnotesize $O$};
    \draw[fill=black!10] (5.4,0) rectangle +(0.2,0.5) node[pos=.5] {\footnotesize $O$};
    \draw[fill=black!10] (5.6,0) rectangle +(0.2,0.5) node[pos=.5] {\footnotesize $O$};
    \draw[fill=black!10] (5.8,0) rectangle +(0.2,0.5) node[pos=.5] {\footnotesize $O$};
    \draw[fill=black!10] (6.0,0) rectangle +(0.2,0.5) node[pos=.5] {\footnotesize $O$};
    \draw[fill=black!10] (6.2,0) rectangle +(0.2,0.5) node[pos=.5] {\footnotesize $O$};
    \draw[fill=black!10] (6.4,0) rectangle +(0.2,0.5) node[pos=.5] {\footnotesize $P$};
    \draw[fill=black!10] (6.6,0) rectangle +(0.2,0.5) node[pos=.5] {\footnotesize $P$};
    \draw[fill=black!10] (6.8,0) rectangle +(0.2,0.5) node[pos=.5] {\footnotesize $P$};
    \draw[fill=black!10] (7.0,0) rectangle +(0.2,0.5) node[pos=.5] {\footnotesize $P$};
    \draw[fill=black!10] (7.2,0) rectangle +(0.2,0.5) node[pos=.5] {\footnotesize $P$};
    \draw[fill=black!10] (7.4,0) rectangle +(0.2,0.5) node[pos=.5] {\footnotesize $P$};
    \draw[fill=black!10] (7.6,0) rectangle +(0.2,0.5) node[pos=.5] {\footnotesize $P$};
    \draw[fill=black!10] (7.8,0) rectangle +(0.2,0.5) node[pos=.5] {\footnotesize $P$};
\draw (4,-0.25) node[transform shape,scale=1.42] {Row in bank 4};
    \draw[fill=green!50!black!30,green!50!black!30,in=270,out=90] (3.85,0) to node{} +(0.05,0.125) to node[rotate=90,yshift=-0.13cm]{
    } +(-0.1,0.25) to node{} +(+0.05,0.125) to node{} +(+0.28,0) to node{} +(-0.05,-0.125) to node{} +(+0.1,-0.25) to node{} +(-0.05,-0.125) to node{} (4.15,0);
    \draw[very thick,green!50!black!30] (3.86,0) -- (4.14,0);
    \draw[very thick,green!50!black!30] (3.86,0.5) -- (4.14,0.5);
    \draw[in=270,out=90] (3.85,0) to node{} +(0.05,0.125) to node[rotate=90,yshift=-0.13cm]{
    } +(-0.1,0.25) to node{} +(+0.05,0.125);
    \draw[in=270,out=90] (4.15,0) to node{} +(0.05,0.125) to node{} +(-0.1,0.25) to node{} +(+0.05,0.125);

  \end{scope}
  \end{tikzpicture}
\caption{Mapping between a 4\,KB page and an 8\,KB DRAM row in the Haswell-EP setup. Banks are numbered $0-7$, pages are numbered $A-P$. Every eighth 64-byte region of a 4\,KB page maps to the same bank in DRAM. In total 8 out of 64 regions ($=512\,B$) map to the same bank. Thus, the memory of each row is divided among 16 different pages ($A-P$) that use memory from the same row. Occupying one of the pages $B-P$ is sufficient to spy on the eight 64-byte regions of page $A$ in the same bank.}
\label{fig:row_slicing}
\end{figure}
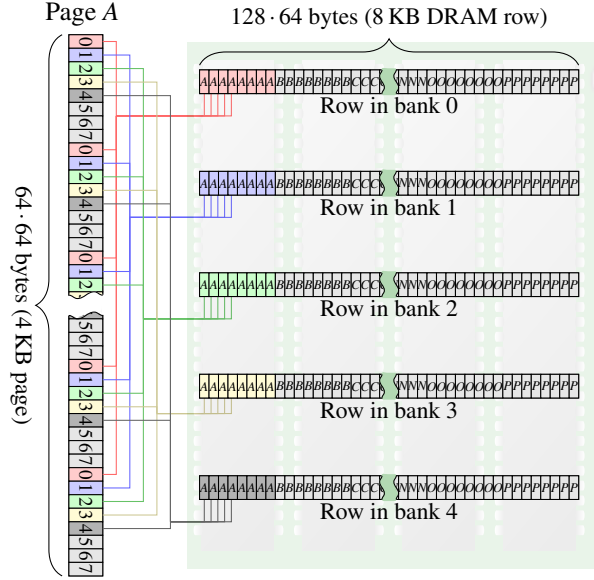

Depending on the addressing functions, a single 4\,KB page can map to multiple DRAM rows. As illustrated in Figure~\ref{fig:row_slicing}, in our Haswell-EP system the contents of a page are split over 8 DRAM rows (with the same row index, but different bank address).
Conversely, a DRAM row contains content of at least two 4\,KB pages, as the typical row size is 8\,KB. More specifically, in our Haswell-EP setup a single row stores content for 16 different 4\,KB pages, as again shown in Figure~\ref{fig:row_slicing}. The amount of memory mapping from one page to one specific row, \eg 512 bytes in the previous case, is the achievable spatial accuracy of our attack.
If none of the DRAM addressing functions uses low address bits ($a_0-a_{11}$), the spatial accuracy is 4\,KB, which is the worst case.
However, if DRAM addressing functions (channel, BG0, CPU, etc.) use low address bits, a better accuracy can be achieved, such as the 512\,B for the server setup. 
On systems where 6 or more low address bits are used, the spatial accuracy of the attack is 64\,B and thus as accurate as a Flush+Reload cache side-channel attack.

Assuming that an attacker occupies at least one other 4\,KB page that maps (in part) to the same bank and row, the attacker has established a situation as illustrated in Figure~\ref{fig:row_hit_attack}. 

To run the side-channel attack on a private memory address $t$ in a victim process, the attacker allocates a memory address $p$ that maps to the same bank and the same row as the target address $t$. As shown in Figure~\ref{fig:row_slicing}, although $t$ and $p$ map to the same DRAM row, they belong to different 4\,KB pages (\ie no shared memory). The attacker also allocates a row conflict address $\bar{p}$ that maps to the same bank but to a different row.

The side-channel attack then works in three steps:
\begin{compactenum}
\item Access the row conflict address $\bar{p}$
\item Wait for the victim to compute
\item Measure the access time on the targeted address $p$
\end{compactenum}
If the measured timing is below a row-hit threshold (cf.\ the highlighted ``row hit'' region in Figure~\ref{fig:hist_cache_comparison}), the victim has just accessed $t$ or another address in the target row.
Thus, we can accurately determine when a specific non-shared memory location is accessed by a process running on another core or CPU.
As $p$ and $\bar{p}$ are on separate private 4\,KB pages, they will not be prefetched and we can measure row hits without any false positives. By allocating all but one of the pages that map to a row, the attacker maximizes the spatial accuracy.

Based on this attack principle, we build a fully automated template attack~\cite{Gruss2015Usenix} that triggers an event in the victim process running on the other core or CPU (\eg by sending requests to a web interface or triggering user-interface events).
For this attack we do not need to reconstruct the full addressing functions nor determine the exact bank address. Instead, we exploit the timing difference between row hits and row conflicts as shown in Figure~\ref{fig:hist_cache_comparison}.

To perform a DRAMA template attack, the attacker allocates a large fraction of memory, ideally in 4\,KB pages. This ensures that some of the allocated pages are placed in a row together with pages used by the victim. The attacker then profiles the entire allocated memory and records the row-hit ratio for each address.

False positive detections are eliminated by running the profiling phase with different events.
If an address has a high row-hit ratio for a single event, it can be used to monitor that event in the exploitation phase. After such an address has been found, all other remaining memory pages will be released and the exploitation phase is started.

\subsection{Evaluation}
We evaluated the performance of our side-channel attack in several tests. These tests were performed on a dual-core laptop with an Ivy Bridge Intel i5-3230M CPU with 2 Samsung DDR3-1600 dual-rank 4\,GB DIMMs in dual-channel configuration.

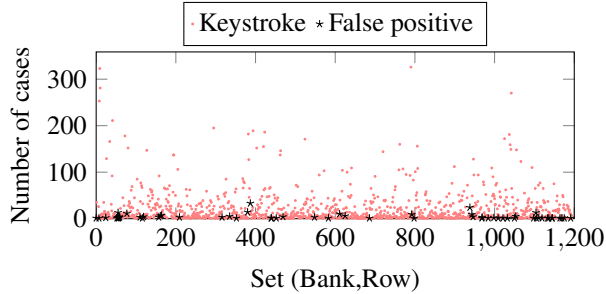
\begin{figure}
  \centering
  \begin{tikzpicture}
\pgfplotsset{compat=newest,every axis legend/.append style={at={(0.5,1.3)},anchor=north}}
  \begin{axis}[
  legend columns=2,
  xlabel={Set (Bank,Row)},
  ylabel={Number of cases},
  width=\hsize,
  ymin=0,
  xmin=0,
  xmax=1200,
  restrict y to domain={1:10000},
  height=3.8cm
]
\addplot+[only marks,mark size=0.33pt,mark=*,solid,red!50,draw=red!50,mark options={draw=red!50,fill=red!50}] table[x=set_row,y=hits_key] {cache_template_ff.csv};
\addlegendentry{Keystroke\,}
\addplot+[only marks,mark size=1.5pt,mark=star,solid,black,draw=black,mark options={draw=black,fill=black}] table[x=set_row,y=hits_nokey] {cache_template_ff.csv};
\addlegendentry{False positive}
\end{axis}
\end{tikzpicture}
  \caption{A DRAM template of the system memory with and without triggering keystrokes in the Firefox address bar. 1136 sets had row hits after a keystroke, 59 sets had false positive row hits (row hits without a keystroke), measured on our Ivy Bridge i5 test system.}
  \label{fig:comparison_firefox}
\end{figure}

The first test was a DRAMA template attack. The attack ran without any shared memory in an unprivileged user program.
In this template attack we profiled access times on a private memory buffer while
triggering keystrokes in the Firefox address bar.
Figure~\ref{fig:comparison_firefox} shows the template attack profile with and without keystrokes being triggered.
While scanning a total of 7\,GB of allocated memory, we found $1195$ addresses that showed at least one row hit during the tests.
$59$ of these addresses had row hits independent of the event (false positives), \ie these $59$ addresses cannot be used to monitor keystroke events. For the remaining $1136$ addresses we only had row hits after triggering a keystroke in the Firefox address bar. Out of these addresses, $360$ addresses had more than 20 row hits. Any of these $360$ addresses can be used to monitor keystrokes reliably. The time to find an exploitable address varies between a few seconds and multiple minutes. Sometimes the profiling phase does not find any exploitable address, for instance if there is no memory in one row with victim memory. In this case the attacker has to restart the profiling phase.

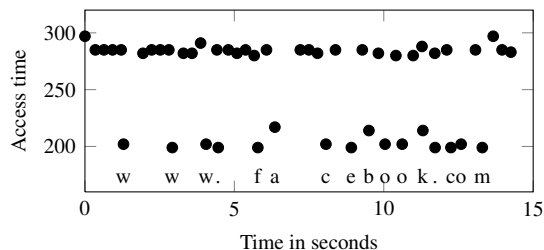
\begin{figure}[t]
\centering
\vspace{-2.2pt}
\begin{tikzpicture}
\begin{axis}[
style={font=\footnotesize},
xlabel=Time in seconds,
ylabel=Access time,
width=0.95\hsize,
xmin=0,
xmax=15,
ymax=320,
ymin=160, 
height=4cm
]
\addplot[only marks] table[x=Time,y=Value] {sentence.log};
\node [text depth=0pt, text height=0.5ex] at (axis cs:  1.2901128867,  170) {\footnotesize{w}};
\node [text depth=0pt, text height=0.5ex] at (axis cs:  2.9231052973,  170) {\footnotesize{w}};
\node [text depth=0pt, text height=0.5ex] at (axis cs:  4.0529646867,  170) {\footnotesize{w}};
\node [text depth=0pt, text height=0.5ex] at (axis cs:  4.4609336536,  170) {\footnotesize{.}};
\node [text depth=0pt, text height=0.5ex] at (axis cs:  5.7856817439,  170) {\footnotesize{f}};
\node [text depth=0pt, text height=0.5ex] at (axis cs:  6.3557472652,  170) {\footnotesize{a}};
\node [text depth=0pt, text height=0.5ex] at (axis cs:  8.0641409279,  170) {\footnotesize{c}};
\node [text depth=0pt, text height=0.5ex] at (axis cs:  8.9144251336,  170) {\footnotesize{e}};
\node [text depth=0pt, text height=0.5ex] at (axis cs:  9.5028393764,  170) {\footnotesize{b}};
\node [text depth=0pt, text height=0.5ex] at (axis cs:  10.0430607000,  170) {\footnotesize{o}};
\node [text depth=0pt, text height=0.5ex] at (axis cs:  10.6170265567,  170) {\footnotesize{o}};
\node [text depth=0pt, text height=0.5ex] at (axis cs:  11.3087122833,  170) {\footnotesize{k}};
\node [text depth=0pt, text height=0.5ex] at (axis cs:  11.7175577658,  170) {\footnotesize{.}};
\node [text depth=0pt, text height=0.5ex] at (axis cs:  12.2471001021,  170) {\footnotesize{c}};
\node [text depth=0pt, text height=0.5ex] at (axis cs:  12.5892906848,  170) {\footnotesize{o}};
\node [text depth=0pt, text height=0.5ex] at (axis cs:  13.2971672630,  170) {\footnotesize{m}};
\end{axis}
\end{tikzpicture}
  \caption{Exploitation phase on non-shared memory in a DRAMA template attack on our Ivy Bridge i5 test system. A low access time is measured when the user presses a key in the Firefox address bar. The typing gaps illustrate the low noise level.}
  \label{fig:plot_firefox}
\end{figure}

After automatically switching to the exploitation phase we are able to monitor the exact timestamp of every keystroke in the address bar. We verified empirically that row hits can be measured on the found addresses after keystrokes by triggering keystrokes by hand. Figure~\ref{fig:plot_firefox} shows an access time trace for an address found in a DRAMA template attack, while typing in the Firefox address bar. For every key the user presses, a low access time is measured.
We found this address after less than 2 seconds. Over 80 seconds we measured no false positive row hits and when pressing 40 keys we measured no false negatives. During this test the system was entirely idle apart from the attack and the user typing in Firefox. In a real attack, noise would introduce false negatives.

\subheading{Comparison with cache template attacks.}
To compare DRAMA template attacks with cache template attacks, we performed two attacks on gedit.
The first uses the result from a cache template attack in a DRAMA exploitation phase.
The second is a modified cache template attack that uses the DRAMA side channel.
Both attacks use shared memory to be able to compare them with cache template attacks.
However, the DRAMA side-channel attack takes no advantage of shared memory in any attack.

In the first attack on gedit, we target tab open and tab close events.
In an experiment over 120 seconds we opened a new tab and closed the new tab, each 50 times.
The exploitable address in the shared library was found in a cache template attack.
We computed the physical address and thus bank and row of the exploitable address using privileged operating services. Then we allocated large arrays to obtain memory that maps to the same row (and bank). This allows us to perform an attack that has only minimal differences to a Flush+Reload attack.

During this attack, our spy tool detected 1 false positive row hit and 1 false negative row hit.
Running \texttt{stress -m 1} in parallel, which allocates and accesses large memory buffers, causes a high number of cache misses, but did not introduce a significant amount of noise. In this experiment the spy tool detected no false positive row hits and 4 false negative row hits. Running \texttt{stress -m 2} in parallel (\ie the attacker's core is under stress) made any measurements impossible. While no false positive detections occurred, only 9 events were correctly detected. Thus, our attack is susceptible to noise especially if the attacker only gets a fraction of CPU time on its core.

\begin{figure}
  \centering
  \begin{tikzpicture}
  \pgfplotsset{compat=newest,every axis legend/.append style={at={(0.5,1.3)},anchor=north}}
  \begin{axis}[
  legend columns=2,
  xlabel={Address},
  ylabel={Number of cases},
  width=\hsize,
  ymin=0,
  restrict y to domain={1:100},
  height=3.8cm
  ]
  \addplot+[only marks,mark size=0.33pt,mark=*,solid,red!50,draw=red!50,mark options={draw=red!50,fill=red!50}] table[x=addr,y=row_hits] {cache_template_gedit.csv};
  \addlegendentry{Row hits\,}
  \addplot+[only marks,mark size=1.5pt,mark=star,solid,black,draw=black,mark options={draw=black,fill=black}] table[x=addr,y=cache_hits] {cache_template_gedit.csv};
  \addlegendentry{Cache hits}
  \end{axis}
  \end{tikzpicture}
  
  \caption{Comparison of a cache hits and row hits over the virtual memory where the \texttt{gedit} binary is mapped, measured on our Ivy Bridge i5 test system.}
  \label{fig:comparison_gedit}
\end{figure}
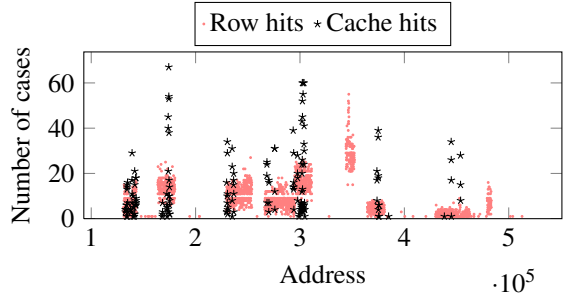

In the second attack we compared the cache side channel and the DRAM side channel in a template attack on keystrokes in gedit.
Figure~\ref{fig:comparison_gedit} shows the number of cache hits and row hits over the virtual memory where the \texttt{gedit} binary is mapped.
Row hits occur in spatial proximity to the cache hits and at shifted offsets due to the DRAM address mappings.

\subsection{Comparison with state of the art}
We now compare DRAMA side-channel attacks with same-CPU cache attacks such as Flush+Reload and Prime+Probe, as well as with cross-CPU cache attacks~\cite{Irazoqui2015cpca}.
Our attack is the first to enable monitoring non-shared memory cross-CPU with a reasonably high spatial accuracy and a timing accuracy that is comparable to Flush+Reload. 
This allows the development of new attacks on programs using dynamically allocated or private memory.

The spatial accuracy of the DRAMA side-channel attack is significantly higher than that of a Prime+Probe attack, which also does not necessitate shared memory, and only slightly lower than that of a Flush+Reload attack in most cases.
Our Ivy Bridge i5 system has 8\,GB DRAM and a 3\,MB L3 cache that is organized in 2 cache slices with each 2048 cache sets. Thus, in a Prime+Probe attack 32768 memory lines map to the same cache set, whereas in our DRAMA side-channel attack, on the same system, only 32 memory lines map to the same row. The spatial accuracy strongly depends on the system. On our Haswell-EP system only 8 memory lines map to the same row whereas still 32768 memory lines map to the same cache set. Thus, on the Haswell-EP system the advantage of DRAMA side-channel attacks over Prime+Probe is even more significant.

To allocate memory lines that are in the same row as victim memory lines, it is necessary to allocate significantly larger memory buffers than in a cache attack like Prime+Probe. This is a clear disadvantage of DRAMA side-channel attacks.
However, DRAMA side-channel attacks have a very low probability of false positive row hit detections, whereas Prime+Probe is highly susceptible to noise. Due to this noise, monitoring singular events using Prime+Probe is extremely difficult.

Irazoqui~\etal\cite{Irazoqui2015cpca} presented cache-based cross-CPU side-channel attacks. However, their work requires shared memory. Our approach works without shared memory. Not only does this allow cross-CPU attacks in highly restricted environments, it also allows to perform a new kind of cross-core attack within one system.

\section{Improving attacks}\label{sec:improving}
In this section, we describe how the DRAM addressing functions can be used to improve the accuracy, efficiency, and success rate of existing attacks.

\subheading{Flush+Reload.} The first step when performing Flush+Reload attacks is to compute a cache-hit threshold, based on a histogram of cache hits and cache misses (memory accesses). However, as we have shown (cf.\ Figure~\ref{fig:hist_cache_comparison}) row hits have a slightly lower access time than row conflicts. To get the best performance in a Flush+Reload attack it is necessary to take row hits and conflicts into account.
Otherwise, if a process accesses any memory location in the same row, a row hit will be misclassified as a cache hit.
This introduces a significant amount of noise as the spatial accuracy of a cache hit is 64 bytes and the one of a row hit can be as low as 8\,KB, depending on how actively the corresponding pages of the row are used. We found that even after a call to \texttt{sched\_yield} a row hit is still observed in $2\%$ of the cases on a Linux system that is mostly idle. In a Flush+Reload attack the victim computes in parallel and thus the probability then is even higher than $2\%$. This introduces a significant amount of noise especially for Flush+Reload attacks on low-frequency events. Thus, the accuracy of Flush+Reload attacks can be improved significantly taking row hits into account for the cache hit threshold computation.

\subheading{Rowhammer.} In a Rowhammer attack, an adversary tries to trigger bit flips in DRAM by provoking a high number of row switches. The success rate and efficiency of this attack benefit greatly from knowing the DRAM mapping, as we now demonstrate.

In order to cause row conflicts, one must alternately access addresses belonging to the same bank, but different row. The probability that 2 random addresses fulfill this criterion is $2^{-B}$, where $B$ is the total number of bank-addressing bits (this includes all bits for channel, rank, etc.). For instance, with the dual-channel DDR4 configuration shown in Figure~\ref{fig:mapping_skylake} this probability is only $2^{-6} = 1/64$. By hammering a larger set of addresses, the probability of having at least two targeting the same bank increases. However, so does the time in between row switches, thus the success rate decreases.

The most efficient way of performing the Rowhammer attack is \emph{double-sided hammering}. Here, one tries to cause bit flips in row $n$ by alternatingly accessing the adjacent rows $n-1$ and $n+1$, which are most likely also adjacent in physical memory.
The most commonly referenced implementation of the Rowhammer attack, by Seaborn and Dullien~\cite{DullienSeaborn2015github}, performs double-sided hammering by making assumptions on, \eg the position of the row-index bits. If these are not met, then their implementation does not find any bit flips. Also, it needs to test multiple address combinations as it does not use knowledge of the DRAM addressing functions. We tested their implementation on a Skylake machine featuring G.SKILL F4-3200C16D-16GTZB DDR4 memory at the highest possible refresh interval, yet even after 4 days of nonstop hammering, we did not detect any bit flips.

By using the DRAM addressing functions we can immediately determine whether two addresses map to the same bank. Also, we can very efficiently search for pairs allowing double-sided hammering. After taking the reverse-engineered addressing functions into account, we successfully caused bit flips on the same Skylake setup within minutes. Running the same attack on a Crucial DDR4-2133 memory module running at the default refresh interval, we observed the first bit flip after 16 seconds and subsequently observed on average one bit flip every 12 seconds.
Although the LPDDR4 standard includes \emph{target row refresh} (TRR) as an optional countermeasure against the Rowhammer attack, the DDR4 standard does not. Still, some manufacturers include it in their products as a non-standard feature. For both DDR4 and LPDDR4, both the memory controller and the DRAM must support this feature in order to provide any protection.
To the best of our knowledge, both our Haswell-EP test system and the Crucial DDR4-2133 memory module, with Micron DRAM chips, support TRR~\cite{MicronDDR4,intelxeonupdate}. However, we are still able to reproducibly trigger bit flips in this configuration.

\section{Countermeasures}\label{sec:countermeasures}
Defending against row buffer attacks is a difficult task. Making the corresponding DRAM operations constant time would introduce unacceptable performance degradation. However, as long as the timing difference exists and can be measured, the side channel cannot be closed.

Our attack implementations use the unprivileged \texttt{clflush} instruction in order to cause a DRAM access with every memory request. Thus, one countermeasure might be to restrict said operation. However, this requires architectural changes and an attacker can still use eviction as a replacement. The additional memory accesses caused by eviction could make our row-buffer covert channel impractical. However, other attacks such as the fully automated reverse engineering or our row-hit side-channel attack are still possible.
Restricting the \texttt{rdtsc} instruction would also not prevent an attack as other timing sources can be used as replacement.

To prevent cross-VM attacks on multi-CPU cloud systems, the cloud provider could schedule each VM on a dedicated physical CPU and only allow access to CPU-local DRAM. This can be achieved by using a non-interleaved NUMA configuration and assigning pages to VMs carefully.
This approach essentially splits a multi-CPU machine into independent single-CPU systems, which leads to a loss of many of its advantages.

Saltaformaggio~\etal\cite{Saltaformaggio2013} presented a countermeasure to the memory bus-based covert channel of Wu \etal. It intercepts atomic instructions that are responsible for this covert channel, so that only cores belonging to the attacker's VM are locked, instead of the whole machine. This countermeasure is not effective against our attacks as they do not rely on atomic instructions.

Finally, our attack could be detected due to the high number of cache misses. However, it is unclear whether it is possible to distinguish our attacks from non-malicious applications.

\section{Conclusion}\label{sec:conclusions}
In this paper, we presented two methods to reverse engineer the mapping of physical memory addresses to DRAM channels, ranks, and banks. One uses physical probing of the memory bus, the other runs entirely in software and is fully automated. 
We ran our method on a wide range of architectures, including desktop, server, and mobile platforms.

Based on the reverse-engineered functions, we demonstrated DRAMA (DRAM addressing) attacks. This novel class of attacks exploits the DRAM row buffer that is a shared resource in single and multi-processor systems. This allows our attacks to work in the most restrictive environments, \ie across processors and without any shared memory. 
We built a covert channel with a capacity of 2\,Mbps, which is three to four orders of magnitude faster than memory-bus-based channels in the same setting.
We demonstrated a side-channel template attack automatically locating and monitoring memory accesses, \eg user input, server requests. This side-channel attack is as accurate as recent cache attacks like Flush+Reload, while requiring no shared memory between the victim and the spy. 
Finally, we show how to use the reverse-engineered DRAM addressing functions to improve existing attacks, such as Flush+Reload and Rowhammer. Our work enables practical Rowhammer attacks on DDR4.

We emphasize the importance of reverse engineering microarchitectural components for security reasons. Before we reverse engineered the DRAM address mapping, the DRAM row buffer was transparent to
operating system and software. Only by reverse engineering we made this shared resource visible
and were able to identify it as a powerful side channel.

\section*{Acknowledgments}
We would like to thank our anonymous reviewers as well as Anders Fogh, Moritz Lipp, and Mark Lanteigne for their valuable comments and suggestions.

Supported by the EU FP7 programme under GA No. 610436 (MATTHEW) and the Austrian Research Promotion Agency (FFG) under grant number 845579 (MEMSEC).

{\footnotesize \bibliographystyle{acm}
\bibliography{bibliography}}

\end{document}